\documentclass[11pt]{article}
\usepackage[margin=1in]{geometry}
\usepackage[T1]{fontenc}
\usepackage[utf8]{inputenc}
\usepackage{textcomp}
\usepackage{amsmath,amssymb}
\usepackage{graphicx}
\usepackage{booktabs,array}
\usepackage{multirow}
\usepackage{enumitem}
\usepackage{caption}
\usepackage{titlesec}
\usepackage[numbers,super,sort&compress]{natbib}
\usepackage[hidelinks]{hyperref}

\makeatletter
\renewcommand\@biblabel[1]{#1.}
\makeatother
\newcommand{\figref}[2][]{\hyperref[#2]{Fig.~\ref*{#2}#1}}
\newcommand{\figsref}[2][]{\hyperref[#2]{Figs.~\ref*{#2}#1}}
\newcommand{\eqnref}[1]{\hyperref[#1]{Eq.~(\ref*{#1})}}
\newcommand{\eqnsref}[2]{\hyperref[#1]{Eqs.~(\ref*{#1})}--\hyperref[#2]{(\ref*{#2})}}
\newcommand{\tabref}[1]{\hyperref[#1]{Table~\ref*{#1}}}

\titlespacing*{\section}{0pt}{0.9em}{0.6em}
\titlespacing*{\subsection}{0pt}{0.8em}{0.25em}
\begin{document}

\begin{center}
{\LARGE\bfseries Self-similarity of mobility networks\par}
\vspace{0.75em}
Ying-Yue Lyu\textsuperscript{1,2,3}, Xiao-Yong Yan\textsuperscript{1,2}*, Bin Jia\textsuperscript{1,2}, Jobst Heitzig\textsuperscript{3}, Ziyou Gao\textsuperscript{1,2}*, and J{\"u}rgen Kurths\textsuperscript{3,4}*

\emph{\textsuperscript{1} School of Systems Science, Beijing Jiaotong University, Beijing 100044, P. R. China.}

\emph{\textsuperscript{2} Hebei Key Laboratory of Future Urban Intelligent Traffic Management, Beijing Jiaotong University, Beijing, 100044, P. R. China.}

\emph{\textsuperscript{3} Potsdam Institute for Climate Impact Research, 14412 Potsdam, Germany.}

\emph{\textsuperscript{4} Department of Physics, Humboldt University of Berlin, 12489 Berlin, Germany.}
\end{center}

\section{Abstract}\label{abstract}

Mobility systems of people and goods are inherently multi-scale, spanning levels of organization from individual cities to regions and nations. Understanding whether mobility networks exhibit similar patterns across these scales is important. Such similarity would point to common organizing principles, enabling insights gained at one scale to inform planning and management at others. Despite growing efforts to analyze mobility at multiple scales, such cross-scale similarity remains poorly understood, and renormalization provides a natural framework for addressing this question. Here, we propose a Neighbor-Limited Box Covering method to renormalize undirected weighted mobility networks. This method iteratively selects box centers in descending order of node strength, merges each center with a fixed number of its highest-weight neighbors to form a renormalized node, and aggregates edge weights between renormalized nodes to generate the network at the next scale. We apply this technique to uncover multi-scale structures of real-world inter-city human mobility and freight trip networks in China and find that the topological structures, weighted structural features, and dynamic processes all exhibit self-similarity across these multi-scale mobility networks. Moreover, we find that the constituent nodes in most renormalized nodes show a strong spatial cohesion, and the boundaries of them closely follow existing political and socio-economic borders, even though the method does not explicitly incorporate any spatial information. Our study not only reveals the consistency of multi-scale inter-city mobility patterns, but also provides important insights into their spatial organization. Furthermore, our method is applicable to mobility networks of different sizes and has potential as a powerful tool for the multi-scale analysis of various other real-world complex systems.

\section{Introduction}\label{introduction}

Mobility, encompassing the movement of both people and goods, underpins socio-economic interactions across cities, regions and nations. Mobility patterns reveal the functional connections among these places and shape urban development, regional integration, and economic activity\cite{ref1}. To understand these patterns, mobility systems are often analyzed through the lens of complex networks\cite{ref2,ref3}, where traffic analysis zones (TAZ) are represented as nodes and flows between them as weighted edges\cite{ref4,ref5,ref6}, thereby forming weighted mobility networks. Traditional research on such networks primarily focuses on pre-defined TAZs, such as grid-based regions or administrative boundaries (e.g., city, province, or country), leading to the analysis of mobility patterns at a single scale. However, real-world movements of people and goods span multiple scales\cite{ref7}, ranging from intra-city and inter-city to regional and international travel\cite{ref8}. This multi-scale nature matters because different scales capture distinct aspects of mobility systems: city-scale mobility reflects commuting structures and daily activity patterns; regional-scale mobility reflects industrial linkages and economic interactions; and international-scale mobility reflects trade flows and global supply chain networks. A comprehensive understanding of mobility systems therefore requires integrating information across multiple scales.

To this end, several multi-scale community detection methods have been employed to reveal hierarchical community structures of mobility networks\cite{ref8,ref9,ref10,ref11,ref12,ref13}. Combining nodes within a single community leads to a mobility network at a larger scale\cite{ref13}. These methods predominantly rely on the topological and weighted structures without considering spatial factors. Nonetheless, most studies reveal that the multi-scale communities are spatially nested, and that, at each scale, nodes within one community tend to be spatially cohesive. While these studies shed light on the spatial organization of multi-scale communities, there remains limited understanding of possible self-similarity properties of these multi-scale mobility networks. Such self-similarity would imply that regularities established at one scale may generalize across others, providing a foundation for cross-scale understanding of mobility networks.

In the context of complex networks, self-similarity analyses commonly employ renormalization group methods. These approaches can be classified into three categories: shortest-path-distance renormalization, degree-thresholding renormalization, and geometric renormalization\cite{ref14}.

\begin{enumerate}
\def\labelenumi{(\roman{enumi})}
\item
  In shortest-path-distance renormalization, the box-covering method\cite{ref15} is the most foundational and widely used. It first identifies the minimum number of non-overlapping boxes to tile the network, where each box contains a set of nodes whose shortest path distance is below a length threshold \(l_{B}\). Then, nodes within one box are merged into a single renormalized node, and connections between two boxes are preserved as one link between two renormalized nodes. Repeated application of this method generates a sequence of renormalized networks. The fractal dimension is then calculated by analyzing the relationship between the number of nodes at different renormalization layers and the corresponding box length \(l_{B}\), thereby quantifying the network's self-similarity. This method has influenced a great number of subsequent studies, such as network evolution\cite{ref16}, universality classes\cite{ref17,ref18}, transport phenomena\cite{ref19,ref20}, and fractal dimensions of real-world networks\cite{ref21,ref22,ref23,ref24}. On the other hand, since finding the minimum number of boxes to cover the network is NP-hard\cite{ref15}, numerous studies have focused on enhancing the computational efficiency of box-covering algorithms\cite{ref25}. A widely used algorithm is the greedy coloring algorithm\cite{ref26}, which transforms the original network into a dual network by connecting nodes with path lengths exceeding the threshold \(l_{B}\). Graph coloring is then applied on this dual network, where nodes sharing the same color are grouped into the same box. In addition, Villegas et al.\cite{ref27} proposed a Laplacian renormalization group method, which is able to identify proper spatiotemporal scales in heterogeneous networks. However, these network covering methods are restricted to unweighted networks. To solve this problem, Wei et al.\cite{ref28} introduced a box-covering algorithm for weighted networks (BCANw) to compute their fractal dimensions.
\item
  The degree-thresholding renormalization\cite{ref29} is based on a hidden metric space embedding, in which nodes are assigned latent coordinates and connections are modeled probabilistically as a function of their pairwise distances in the hidden space, thereby providing a geometric representation of the network structure. Renormalization is then performed by removing nodes, whose degrees fall below a given threshold. Increasing this threshold progressively generates a sequence of coarse-grained subgraphs with self-similar structure. This approach has also been applied to various systems, including bipartite networks\cite{ref30}, renormalization flows\cite{ref31}, and self-similarity of email networks\cite{ref32}.
\item
  The geometric renormalization\cite{ref33} is also based on the same hidden metric space embedding. Instead of removing nodes, it merges adjacent nodes in the hidden metric space into super-nodes, and aggregates connections between different super-nodes into single links, thereby forming a renormalized network. Its repeated application also yields a sequence of renormalized networks with self-similarity. Zheng et al.\cite{ref34} extended this approach into the geometric branching algorithm to model network evolution. For weighted networks, Chen et al.\cite{ref35} proposed a sum-weighted geometric renormalization method, which first applies geometric renormalization on an unweighted network and then assigns weights to the renormalized network by calculating the inter-super-node weights as the sum of their shared-link-weights. Zheng et al.\cite{ref36} further refined this by introducing a sup-weighted geometric renormalization method, where inter-super-node weights are defined as the maximum of their shared-link-weights. Both sum- and sup-weighted geometric renormalization methods analyze the distribution function of weighted structural features to examine the self-similarity of renormalized networks.
\end{enumerate}

Despite significant advancements in the development and application of network renormalization methods over the past two decades, applying these techniques to mobility networks entails several challenges. First, assuming that edges and weights in mobility networks represent real-world interactions, it is essential to adopt renormalization methods specifically designed for weighted networks. The sum- and sup-weighted geometric renormalization methods can produce renormalized weighted networks; however, due to the redefinition of connections, mapping a network into a hidden metric space introduces discrepancies in edges and weights between the original network and its hidden metric representation. These discrepancies hinder the accurate multi-scale structure representation of mobility networks. Additionally, the degree-thresholding renormalization method yields a renormalized network by removing nodes, and fails to capture the complete multi-scale mobility patterns. Recently, Jiang et al.\cite{ref37} used the BCANw algorithm to calculate the fractal dimension of human mobility networks. The BCANw algorithm, however, has limitations in characterizing the weighted structure of multi-scale mobility networks. To address this issue, they subsequently proposed a fixed-number renormalization group method, in which nodes are first ranked in descending order of strength, and a fixed number of nodes are then assigned to each box accordingly\cite{ref37}. While intuitively, this strategy ignores network topology and fails to ensure direct connectivity among nodes within the same box. Such a shortcoming is particularly important for mobility networks, where connectivity and its associated weights capture the essential interaction structure and its intensity. However, a renormalization method for mobility networks that preserves both connectivity and its associated edge weight properties is still lacking.

To overcome these difficulties, in this paper, we propose a Neighbor-Limited Box Covering method for renormalizing weighted mobility networks. This method uses critical interactions to group a limited number of neighboring nodes into the same box, thereby preserving both connectivity and its associated edge weight properties of the original mobility networks. We then apply this method to uncover the multi-scale structures of the inter-city human mobility and the inter-city freight trip network in China. We further calculate the fractal dimensions, weighted structural features, and dynamic processes to examine the similarities of these multi-scale networks. We mainly find that both networks exhibit self-similarity at multiple scales. Additionally, we plot the multiple scales of the mobility networks on corresponding geographic maps. To our surprise, despite the renormalization progress being entirely independent of spatial factors, nodes grouped into the same box at each scale exhibit spatial cohesion. Our analysis of these findings provides new insights into the multi-scale organization of mobility systems and advances our understanding of the mechanisms underlying the distinct spatial organization of human and freight mobility.

\section{Results}\label{results}

\subsection{Neighbor-Limited Box Covering (NLBC) Method}\label{neighbor-limited-box-covering-nlbc-method}

The NLBC method renormalizes undirected, weighted networks, with mobility networks as a representative application. In such networks, nodes represent zones of comparable importance, edges represent mobility connections between zones, and edge weights measure the total mobility flow along each connection (cf. Methods for details).

To preserve the connectivity and edge weight properties of mobility networks during renormalization, NLBC follows three criteria. First, the dominant mobility structure should be preserved throughout renormalization. Accordingly, nodes associated with intense mobility flows are preferentially selected as box centers and grouped with their highest-weight neighbors. Second, the maximum number of nodes permitted within a box defines the renormalization scale, and at each scale the number of boxes should be minimized while ensuring complete network coverage. Because finding the minimum box covering is NP-hard\cite{ref15}, NLBC approximates this objective with a greedy strategy that constructs large boxes whenever possible. Finally, interactions between boxes should be faithfully preserved across renormalization scales. To this end, edge weights between renormalized nodes are defined by aggregating all mobility flows between their corresponding boxes. Guided by these criteria, NLBC iteratively builds a hierarchy of renormalized networks through a layer-by-layer box-covering process (\figsref[a--b]{fig:nlbc-method}).

We define the maximum number of nodes permitted within a box as the box mass \(m\). Successive renormalized networks are indexed by the renormalization layer \(l\). Specifically, starting from a prescribed box mass \(m\), NLBC first ranks nodes in descending order of node weight to select box centers (cf. \eqnref{eq:node-weight} in Methods), and groups each center with its \(m - 1\) strongest neighbors into a box of mass \(m\). Once no node satisfies the current neighborhood requirement, this requirement is relaxed step by step to allow smaller boxes to form, and this process repeats until every node has been assigned to a box. The node weights and edge weights of the renormalized network are then obtained by summing the corresponding original weights, thereby preserving cumulative mobility flows across successive renormalization layers. Formal definitions of the network measures and a complete description of NLBC algorithm are provided in Methods.

\subsection{Application of NLBC to real-world mobility networks}\label{application-of-nlbc-to-real-world-mobility-networks}

We construct inter-city human mobility and freight trip network as undirected, weighted mobility networks (\figsref[c-d]{fig:nlbc-method}). The underlying data are obtained from two real-world sources: the Sina Weibo check-in records and the GPS trajectory data from the China Road Freight Supervision and Service Platform, respectively (cf. Methods). To characterize their multi-scale structure, we consider two schemes of renormalizations. The first is a \emph{layer-by-layer} renormalization (LL-Renorm) with a fixed box mass. Taking \(m = 2\) as an example, this process yields a sequence of renormalized networks through successive application of the NLBC method. In LL-Renorm, the observation scale increases with the number of renormalization layers. The second is a \emph{single-layer} renormalization (SL-Renorm) with a variable box mass. In this case, only a single renormalization is applied to the original network, while the box mass is progressively increased to produce a sequence of renormalized networks. In SL-Renorm, the observation scale increases progressively with the box mass (see \(l = 1\) of \figsref[a-b]{fig:nlbc-method}). We apply both LL-Renorm and SL-Renorm to the inter-city human mobility and freight trip networks to uncover their multi-scale structures.

\begin{figure}[htbp]
\centering
\makebox[\linewidth][c]{%
  \includegraphics[width=0.65\linewidth]{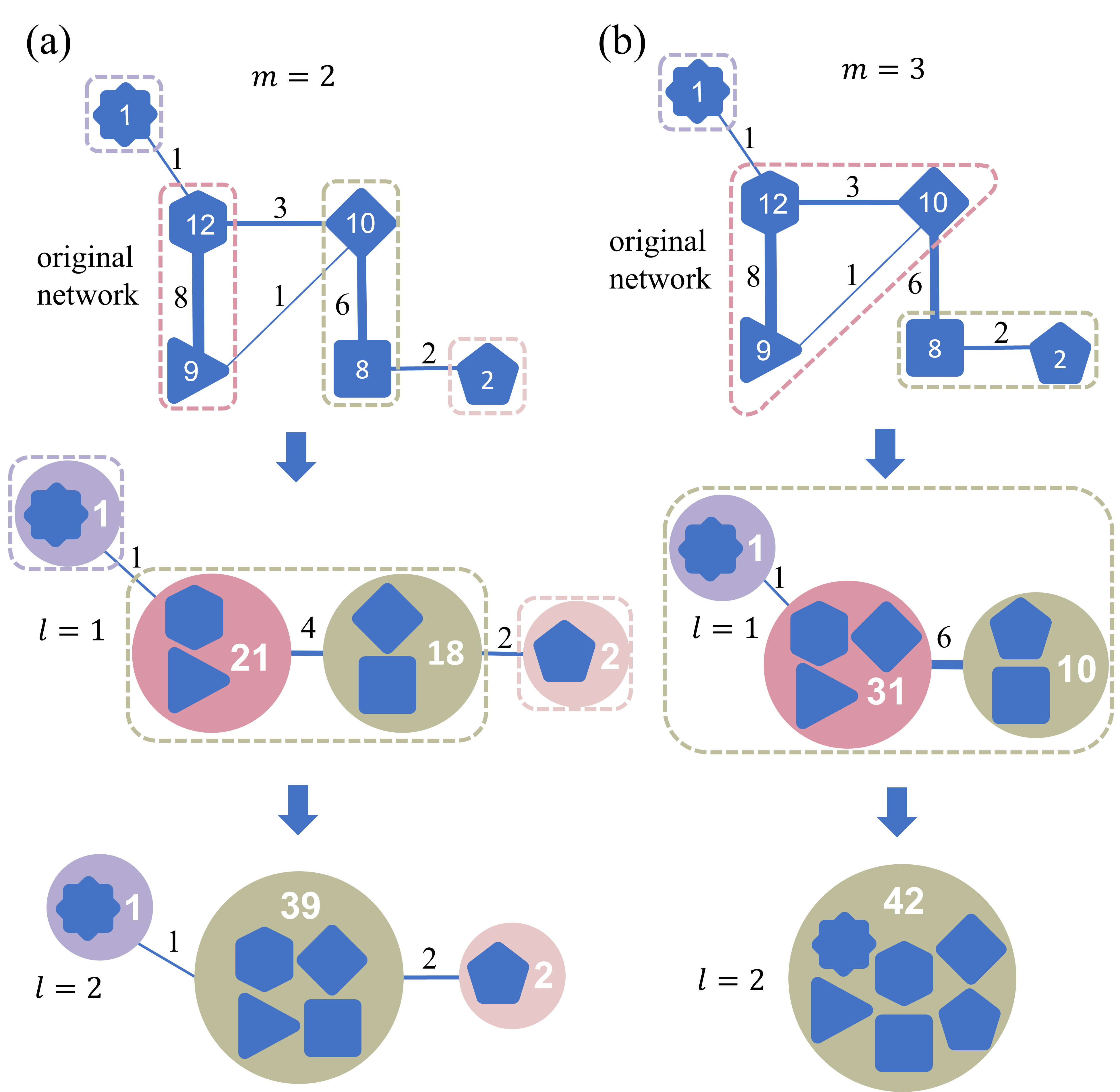}%
  \hspace{0.01\linewidth}%
  \includegraphics[width=0.33\linewidth, trim=20 14 20 10, clip]{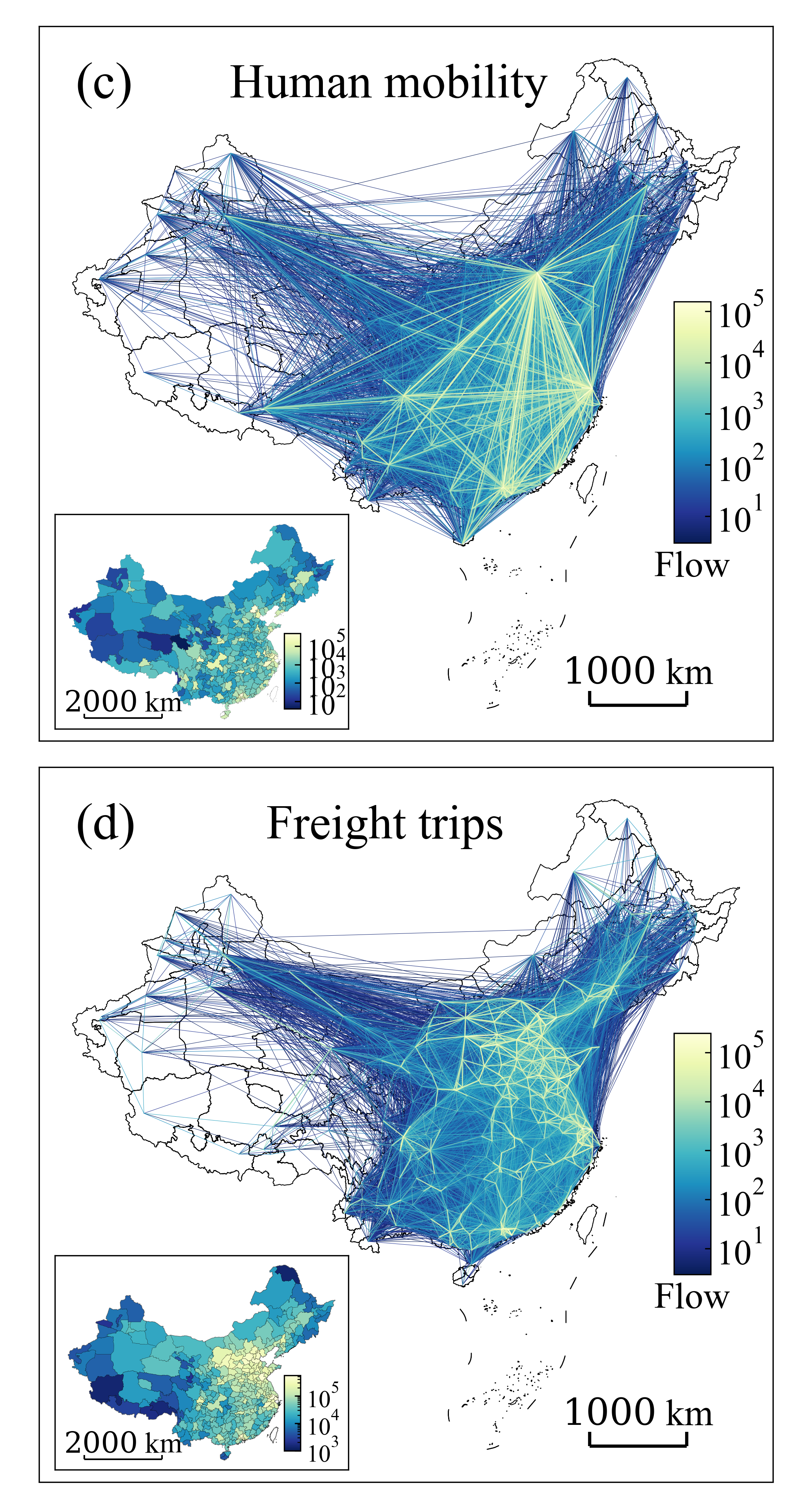}%
}
\caption{\textbf{The NLBC method and the real-world mobility networks.} (a, b) Schematic illustration of the NLBC renormalization process for (a)\(m = 2\) and (b)\(m = 3\), where \(m\) denotes the maximum number of nodes inside one box. Importantly, the node weight differs from node strength. In the original network, the node weight refers to the node strength (see node values of initial networks in a-b); in the renormalized networks, the node weight refers to the sum of the strengths of all original nodes contained within each renormalized node (see the values in the large circles of \(l = 1,2\) in a-b). E.g., in the second layer (\(l = 2\)) of (a), node weight of the central node is 39, whereas its node strength, calculated as the sum of the weights of its two incident edges, is 3. (c) The inter-city human mobility network and (d) the inter-city freight trip network in China.}
\label{fig:nlbc-method}
\end{figure}

\subsection{Self-similarity of multi-scale mobility networks.}\label{self-similarity-of-multi-scale-mobility-networks.}

\begin{figure}[htbp]
\centering
\includegraphics[width=\linewidth]{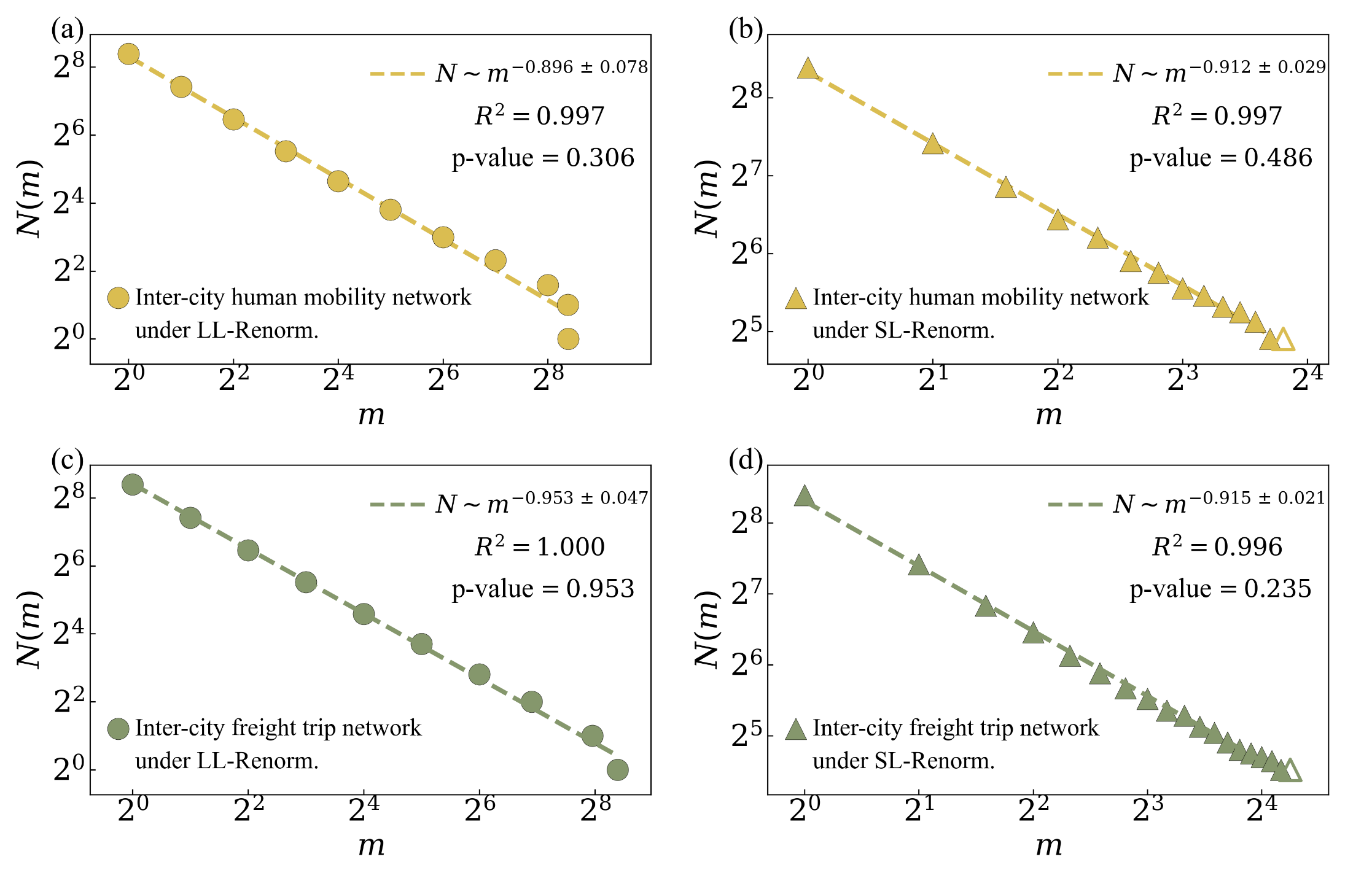}
\caption{\textbf{Self-similar scaling in mobility networks.} Panels (a) and (b) show the inter-city human mobility network under LL-Renorm and SL-Renorm, respectively, while panels (c) and (d) present the inter-city freight trip network under the same renormalization schemes. The LL-Renorm process terminates when only a single node remains. In contrast, the SL-Renorm process terminates when the box mass increases, while the total number of boxes remains constant, as indicated by the hollow triangles in panels (b) and (d). The scaling exponent between \(N\) and \(m\) gives the fractal dimension \(d_{f}\).}
\label{fig:self-similar-scaling}
\end{figure}

We first investigate the self-similar scaling of the multi-scale mobility networks by estimating the fractal dimension \(d_{f}\) following \eqnref{eq:fractal-dimension} in Methods. As shown in \figref{fig:self-similar-scaling}, the box number \(N(m)\) scales with box mass \emph{m} as a power law for human and freight mobility networks under two renormalization schemes, yielding fractal dimensions ranging from 0.896 to 0.953. To verify whether the distributions statistically support a power-law model rather than merely appearing linear in the log-log plot, we perform statistical tests\cite{ref38} (cf. Methods). \figref{fig:self-similar-scaling} shows that all p-values exceed 0.1, indicating that the power-law hypothesis cannot be rejected. Accordingly, both networks exhibit pronounced self-similarity across scales. These findings suggest that, despite the inherent randomness and diversity in individual mobility behaviors, overall mobility patterns are consistent and stable across scales, highlighting their scale-free nature.

\begin{figure}[htbp]
\centering
\includegraphics[width=\linewidth]{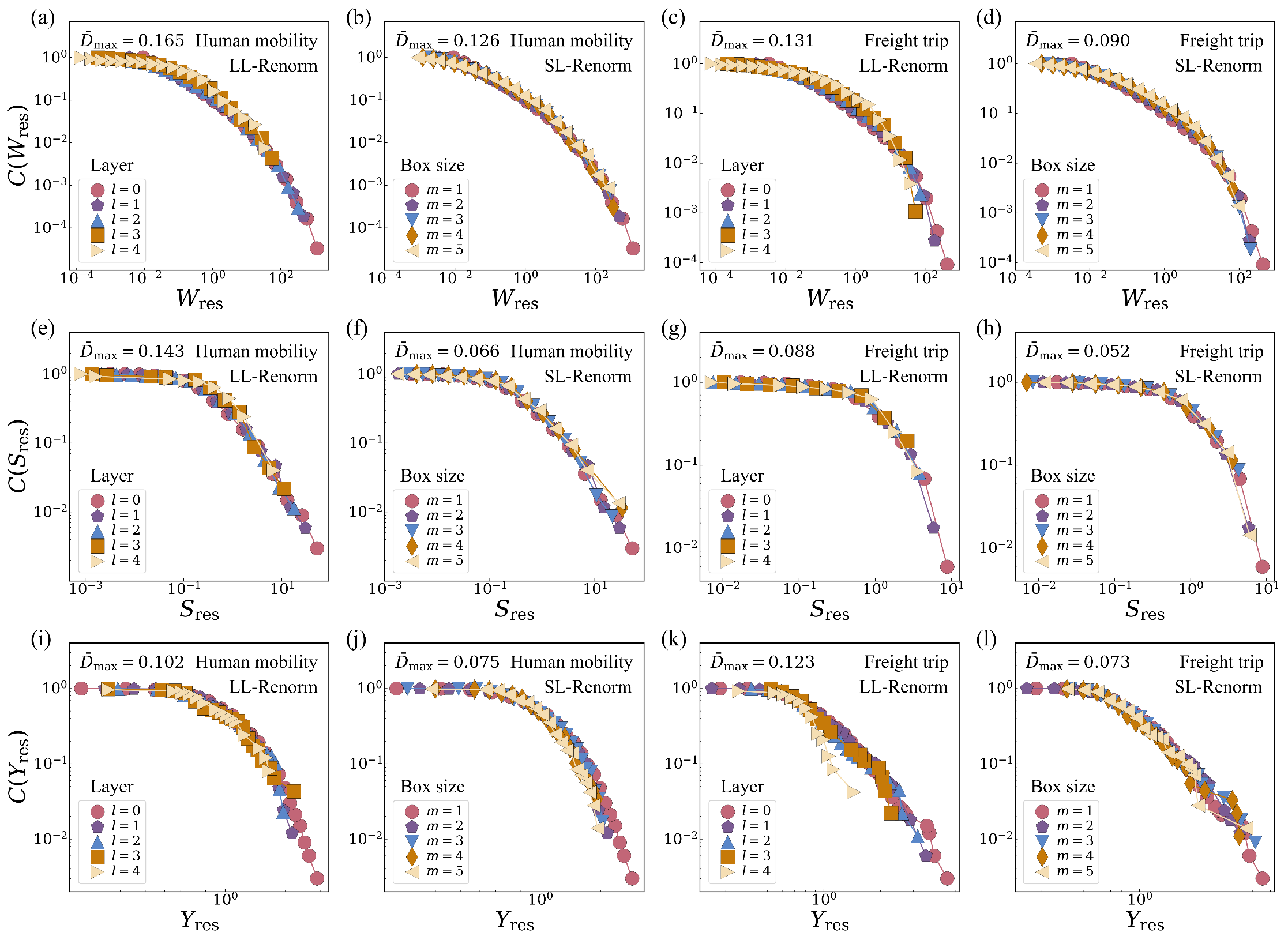}
\caption{\textbf{The weighted structural features for the multi-scale mobility networks.} The complementary cumulative distribution functions of edge weights (a-d), node strengths (e-h) and node disparities (i-l). We apply log-binning to the data and rescale the edge weights, node strengths, and node disparities by the average values of the corresponding renormalized networks.}
\label{fig:weighted-structural-features}
\end{figure}

Although fractal dimensions reveal topological self-similarity in mobility networks, topology alone does not fully capture the weighted structure of mobility systems. We further show that weighted structural features remain invariant across renormalization scales. Specifically, we analyze several indicators, including edge weight \(w_{ij}\), node strength \(S_{i}\), and node disparity \(Y_{i}\) (cf. \eqnsref{eq:node-strength}{eq:node-disparity} in Methods). The complementary cumulative distribution functions (CCDFs) of edge weights (\figref[a-d]{fig:weighted-structural-features}), node strengths (\figref[e-f]{fig:weighted-structural-features}), and node disparities (\figref[i-l]{fig:weighted-structural-features}) merge onto nearly identical curves after rescaling by their respective average values for each layer. To further quantify the degree of similarity among the CCDFs in each group, we compute the average maximum pointwise deviation \({\overline{D}}_{\max}\) (cf. \eqnref{eq:max-pointwise-deviation} in Methods). As shown in \figref{fig:weighted-structural-features}, the values of \({\overline{D}}_{\max}\)are below 0.2 across all subfigures, indicating that the weighted structural features of both inter-city human mobility and freight trip network exhibit statistical self-similarity under both LL-Renorm and SL-Renorm transformations.

From a mobility perspective, the observed self-similarity in node strength suggests that a small number of nodes consistently maintain high strength across scales, representing core nodes that carry the majority of the mobility flow and exert a dominant influence at multiple scales. Similarly, the self-similarity of edge weights reveals that only a few edges maintain high weights across all scales, highlighting critical interactions that dominate mobility flows between nodes, regardless of the observation scale. Furthermore, the self-similarity of node heterogeneity suggests that heterogeneity of interactions between nodes and their neighbors is preserved under renormalization. These findings indicate that the typical characteristics for both human mobility and freight trips follow consistent patterns across scales.

Given the self-similarity observed in both the topological and weighted structures of the multi-scale mobility networks, it is natural to ask whether such properties extend to dynamical processes. Since human mobility forms the backbone of contact patterns through which infectious diseases spread\cite{ref39}, epidemic dynamics on mobility networks provide a natural framework for investigating this question. We therefore simulate epidemic spreading processes on multi-scale mobility networks to examine whether they exhibit similar scale-invariant behavior. 

\begin{figure}[htbp]
\centering
\includegraphics[width=\linewidth]{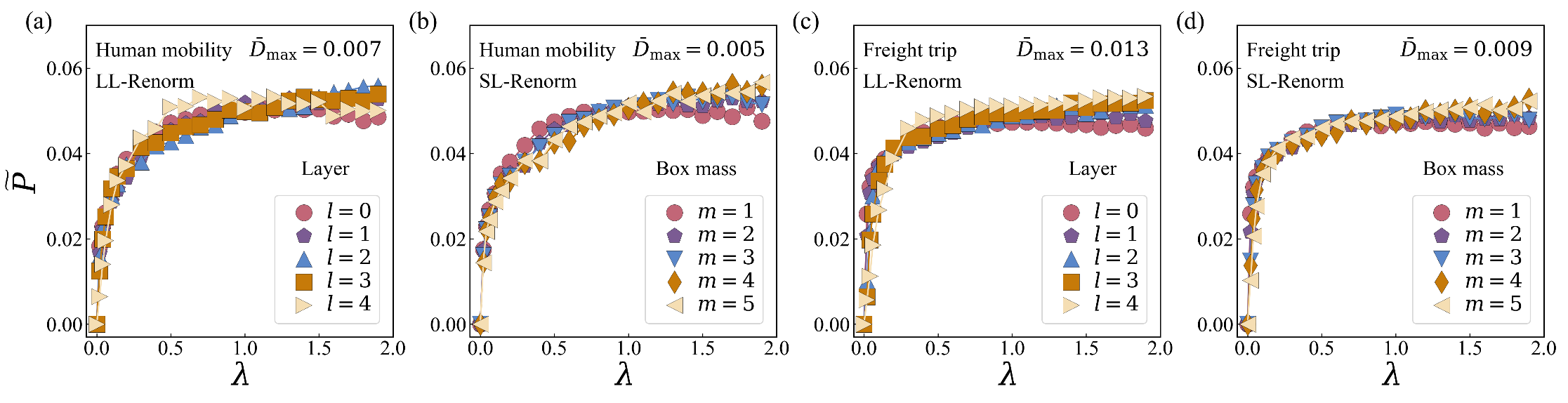}
\caption{\textbf{Epidemic spreading dynamics on multi-scale mobility networks.} The horizontal axis represents the infectivity \(\lambda\). The vertical axis represents the relative infection proportion \(\widetilde{P}.\) If the resulting curves of different renormalized networks nearly overlap, this indicates that networks at different scales exhibit consistent relative infection levels for the same infectivity values.}
\label{fig:epidemic-dynamics}
\end{figure}

Here, we use a weighted susceptible--infected--susceptible (SIS) epidemic spreading model\cite{ref40} (cf. Methods). In this model, epidemic spreading is governed by an infection rate \(\lambda\), which is varied from 0 to 2 using a non-uniform step size, and a recovery probability \(\mu = 1\). Initially, 1\% of nodes are randomly selected as infected, while the remaining nodes susceptible. The epidemic then spreads from infected nodes to their neighbors through weighted edges. Then the spreading process terminates when the infection density reaches a steady state or when a predefined maximum time step is reached. The network size at each renormalization scale is summarized in \tabref{tab:renormalized-network-size} (cf. Methods). For each group of renormalized networks, we compute the relationship between the infectivity \(\lambda\) and the relative infection proportion \(\widetilde{P}\) (cf. \eqnref{eq:relative-infection-proportion} in Methods), as shown in \figref{fig:epidemic-dynamics}, from which we can learn that the curves exhibit similarity across scales. \figref{fig:epidemic-dynamics} also shows that the average maximum pointwise deviations \({\overline{D}}_{\max}\) are low for all the groups, indicating that the dynamical processes on the multi-scale mobility networks remain consistent across scales. As spreading processes are governed by both network topology and weighted structure, this similarity further confirms the cross-scale consistency of these properties, thereby explaining the observed similarity in dynamical behavior.

\subsection{Spatial organization of multi-scale mobility networks}\label{spatial-organization-of-multi-scale-mobility-networks}

\begin{figure}[!htbp]
\centering
\includegraphics[width=0.98\linewidth,height=0.90\textheight,keepaspectratio]{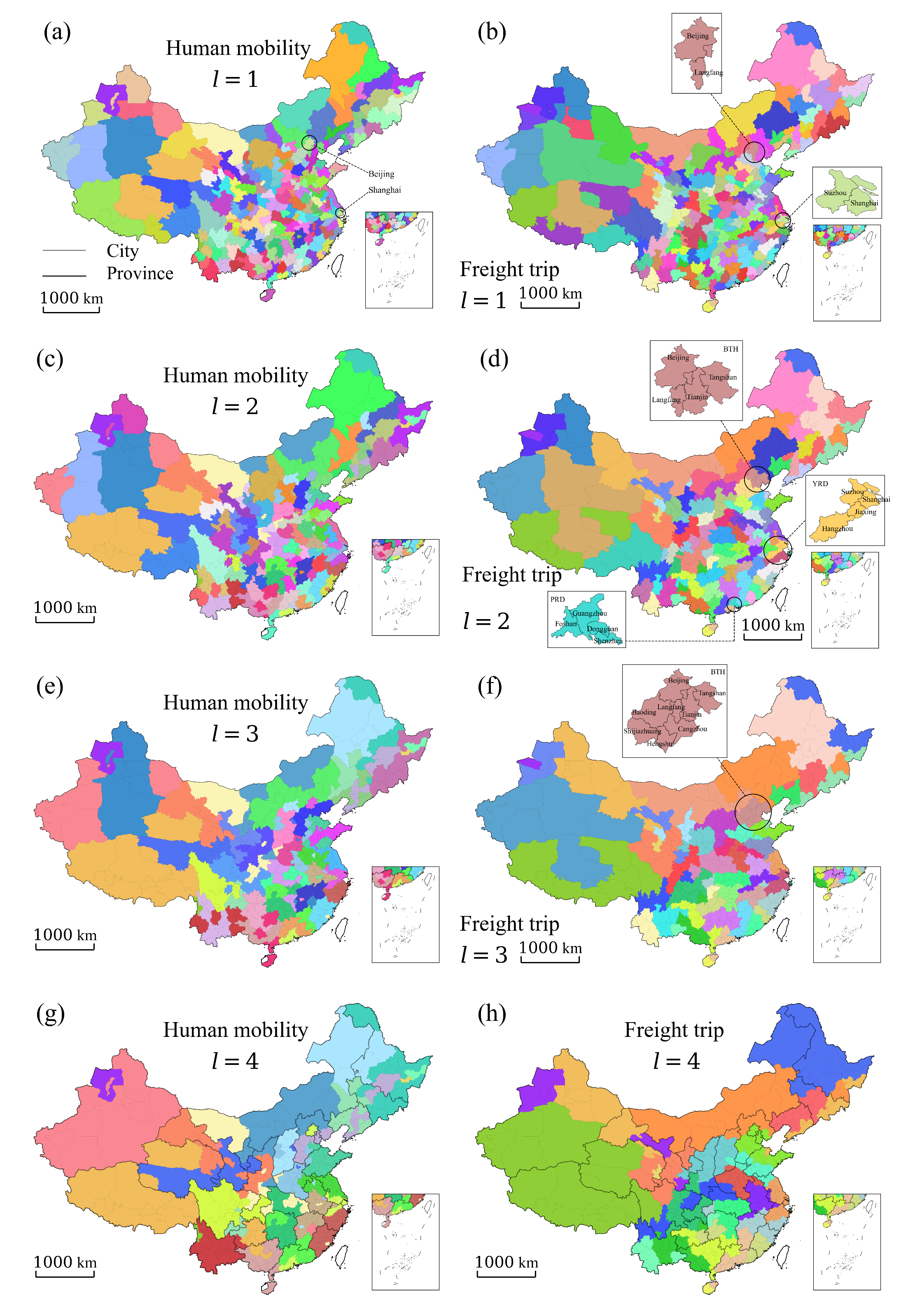}
\caption{\textbf{Geographic maps of the first four LL-Renorm layers for the inter-city human mobility (left) and freight trip (right) networks.} Nodes within the same renormalized node at each scale are represented by the same color. The light lines represent the city-level administrative boundaries, while the bold lines represent the provincial-level administrative boundaries of China.}
\label{fig:geographic-maps}
\end{figure}

The analyses above focus on the structural and dynamical properties of the multi-scale mobility networks. However, mobility systems are fundamentally grounded in geographic space. To complement these analyses, we examine the spatial organization of the multi-scale mobility networks. By plotting the first four layers of the human mobility and freight trip networks under the LL-Renorm (\(m = 2\)) on geographic maps (\figref{fig:geographic-maps}), we observe a clear spatial cohesion among nodes belonging to the same renormalized node.

In the first LL-Renorm layer (\(l = 1\)), both the inter-city human mobility network and the inter-city freight trip network are partitioned into 171 renormalized nodes (\figref[a-b]{fig:geographic-maps}). Most cities within each renormalized node are spatially cohesive, indicating that the geographic distance plays a crucial role in shaping both human and freight mobility. However, differences emerge between both systems:

\begin{enumerate}
\def\labelenumi{(\roman{enumi})}
\item
  In the human mobility network, a small number of distant metropolitan cities with large mobility flows (\figref[a]{fig:nlbc-method}) are aggregated at this step, whereas other cities predominantly merge with geographically proximate neighbors (\figref[a]{fig:geographic-maps}). For instance, although Beijing and Shanghai are geographically distant, they exhibit substantial human mobility flows, leading them to merge in the first step. This suggests that human mobility is shaped not only by distance but also by urban vibrancy and economic development.
\item
  In contrast, freight mobility is far more localized (\figref[b]{fig:nlbc-method}), resulting in most cities merging with nearby neighbors (\figref[b]{fig:geographic-maps}). For example, freight trip flows between Beijing and Shanghai are relatively limited, whereas both cities exhibit larger flows with nearby cities, leading them to merge with their proximal neighbors. This pattern is likely driven by travel costs and freight demand, which favor short-distance movements.
\end{enumerate}

In the second LL-Renorm layer (\(l = 2\)), the inter-city human mobility and freight trip network are partitioned into 88 and 87 renormalized nodes, respectively (\figref[e-f]{fig:geographic-maps}). At this scale, we find that the boundaries of several renormalized nodes in the freight trip network closely align with the core areas of major urban agglomerations in China. As illustrated in \figref[d]{fig:geographic-maps}, the circles delineate the core regions of three such urban agglomerations: the Beijing-Tianjin-Hebei (BTH), the Yangtze River Delta (YRD), and the Pearl River Delta (PRD). By contrast, this alignment is much weaker in the human mobility network, suggesting that the freight trip flows better capture urban agglomeration patterns. These agglomerations, which originated from concentrations of industrial activity, facilitate inter-city collaboration through complementary industrial specializations. As a result, cities within these industrial regions tend to merge into the same renormalized nodes in the freight trip network. For example, within the BTH urban agglomeration (top subfigure in \figref[d]{fig:geographic-maps}), Langfang functions as a logistics and warehousing hub supporting the supply chains of surrounding cities, while Tangshan specializes in heavy industry and construction materials. Tianjin, with its advanced manufacturing and port economy, plays a key role in shipping, logistics, and modern services. Meanwhile, Beijing serves as a major center for finance and trade. This functional division of labor facilitates close collaboration along the industrial chain, thereby promoting coordinated development across the region. In the third LL-Renorm layer (\(l = 3\)) of the freight trip network (\figref[f]{fig:geographic-maps}), the renormalized nodes encompass a broader set of cities within the urban agglomeration. In the BTH, for example, this expansion extends to additional cities in the Hebei Province, such as Shijiazhuang, Baoding, Cangzhou, and Hengshui. Such aggregation pattern highlights the critical role of urban freight transportation in facilitating regional integration.

In the fourth (\(l = 4\)) LL-Renorm layer, the inter-city human mobility and freight trip network are partitioned into 26 and 24 renormalized nodes, respectively (\figref[g-h]{fig:geographic-maps}). At this scale, we find that the boundaries of several renormalized nodes in the human mobility network closely align with provincial administrative boundaries in China. This finding demonstrates that these spatial divisions of human mobility are strongly constrained by socio-administrative boundaries. However, the freight trip network shows less overlap with provincial boundaries. This difference likely stems from the distinct purposes of human and freight mobility. Human mobility is largely driven by individual needs---such as work, education, leisure, and family visits---most of which are closely related to administrative boundaries. However, freight mobility is driven by economic activities and regional trade, which makes it less constrained by provincial boundaries. Together, these results reveal distinct spatial organization patterns across multiple scales between the human mobility and freight trip networks. The mechanisms underlying these patterns are discussed in the Discussion section.

\section{Discussion}\label{discussion}

Previous studies have demonstrated that the multi-scale community structures of human mobility networks exhibit spatial coherence. However, few of them have shown self-similarity properties of mobility networks across multiple scales. Box-covering methods\cite{ref15,ref25} provide a powerful framework for uncovering self-similarity of complex networks. Building on this framework, we propose an NLBC renormalization method for undirected, weighted mobility networks. This technique aggregates highest-weight neighbors into renormalized nodes and constructs new weighted connections based on the structure of the previous layer. Moreover, it preserves mobility interactions within individual TAZs, thereby more closely reflecting the interacting cluster patterns observed in real-world mobility systems. We apply the NLBC method to the inter-city human mobility and freight trip networks in China to uncover their multi-scale structures. Across network scales, we uncover clear self-similarity in topological structures, weighted structural features, and dynamic processes. Owing to its efficiency in generating renormalized replicas of large-size networks, the proposed method is further employed to analyze the multi-scale structures of intra-city human mobility and freight trip networks in four Chinese cities: Beijing, Shanghai, Tianjin, and Chongqing (see Supplementary Materials). The results indicate that the renormalization procedure preserves self-similarity in both the topological and weighted structures of intra-city mobility networks. Collectively, these findings suggest that self-similarity is a universal property inherent in mobility networks.

Beyond theoretical interest, this self-similar property carries important implications. Topological self-similarity suggests that mobility networks share a common backbone structure across scales. For example, mobility patterns identified at city level could be preserved at provincial and regional levels, allowing planning approaches developed at one scale to inform those at others. Weighted self-similarity indicates that key cities and mobility corridors remain prominent across scales, providing a foundation for resilience assessment. Dynamical self-similarity implies that spreading processes can be used to infer dynamics across scales. Local spreading patterns can inform broader-scale dynamics, whereas large-scale patterns help infer spreading processes at finer scales. Together, these implications highlight the practical value of self-similarity for the assessment, prediction, and management of urban and regional systems.

Furthermore, we map the renormalized nodes at each scale onto the geographic space to examine their geospatial organization. Despite the absence of explicit spatial factors in the NLBC process, nodes grouped into most renormalized nodes exhibit a clear spatial cohesion in both inter-city human mobility and freight trip networks. This finding suggests that stronger-weight neighbors tend to be spatially proximate in mobility networks. In human mobility, although long-distance connections are present, short-distance connections remain predominant. In contrast, freight trips are more localized with short-distance connections accounting for a significantly larger proportion. In addition, as the renormalization layer grows, this method, while relying solely on interaction intensities for the node merging, still effectively captures our current understanding of political and socio-economic boundaries. In particular, the fourth-layer renormalized nodes of the inter-city human mobility network align closely with some provincial boundaries, while the second-layer renormalized nodes of the inter-city freight trip network align well with core regions of several urban agglomerations.

In human mobility networks, the alignment between renormalized nodes and provincial administrative boundaries suggests that human mobility behavior appears to be strongly influenced by socio-cultural factors and provincial governance. Over long periods, population migration and cultural diffusion have fostered shared regional identities, including common language, customs, and social norms. These shared backgrounds contribute to the formation of stable social networks and cultural communities, which in turn reinforce mobility interactions among residents. In China, these regional identities have also contributed to the delineation of provincial boundaries. Modern governance further strengthens these interaction patterns, as many routine activities, including education, healthcare, and public services, are organized within provincial jurisdictions. Moreover, transportation infrastructure and services are largely planned and operated within administrative frameworks, strengthening connectivity within provinces while limiting interactions across provincial boundaries. Together, these socio-cultural and institutional mechanisms help generate stronger within-province mobility interactions, resulting in the observed alignment between renormalized nodes and provincial boundaries.

In freight trip networks, by contrast, the alignment of renormalized nodes with urban agglomeration boundaries suggests that freight mobility is primarily driven by freight demand, industrial specialization and logistics efficiency. As a result, freight interactions are structured by supply-chain linkages and inter-city economic complementarities that transcend administrative boundaries. Urban agglomerations therefore provide a more functional spatial unit for capturing these interactions. However, freight governance and infrastructure planning remain partly organized around provincial jurisdictions rather than functional economic regions. This mismatch can contribute to fragmented logistics planning, duplicated infrastructure investment and weaker coordination across provincial borders. Identifying cities with strong freight interactions, particularly those spanning provincial boundaries, therefore provides useful reference points for improving freight infrastructure, promoting regional specialization and supporting more coordinated development within urban agglomerations.

While our approach demonstrates promising performance in transportation applications, a fixed number of boxes is adopted for simplicity. In other domains, however, the NLBC can be further improved by adaptively adjusting the number of boxes through the incorporation of additional influencing factors, thereby better reflecting practical conditions. Such extensions would be based on a more comprehensive assessment of relevant system characteristics, enhancing the accuracy, reliability, and applicability of the results. More generally, the extended NLBC renormalization method can be applied to provide alternative multi-scale descriptions for other complex systems characterized by interaction among nodes. For example, in an international trade network, the extended method could reveal cooperative patterns across multiple scales (e.g., country, regional, and continental levels), helping to identify key trading partners and providing a scientific basis for formulating targeted regional trade policies. Another potential application lies in the multi-scale analysis of scientific collaboration networks. Compared with traditional approaches confined to a single scale, our proposed method allows to uncover collaboration patterns across multiple levels (individual researchers, research teams, and research fields), offering valuable insights for governments and research management institutions to optimize resource allocation and promote interdisciplinary collaboration. Overall, these applications highlight the potential of the NLBC method as a powerful tool for the multi-scale analysis of complex systems, enabling deeper insights into their underlying structures.

\section{Methods}\label{methods}

\subsection{Data description}\label{data-description}

We collect both human mobility data and freight trip data for prefecture-level cities in China. The human mobility data are obtained from Yan et al.\cite{ref7} They extracted individual mobility trajectories across Chinese cities from Sina Weibo check-in records throughout the year 2010. Each trajectory includes attributes such as user ID, check-in time, and check-in location (latitude and longitude). Owing to the upgrade of Danzhou (a city in Hainan Province) to a prefecture-level city in 2015, the Sina Weibo check-in data cover 335 prefecture-level cities, excluding Danzhou.

The freight trip data, obtained from Yang et al.\cite{ref41}, cover 336 prefecture-level cities in China during the period from 18 May 2018 to 31 May 2018. These data include the geographic coordinates of the origin and destination of each trip and are derived from heavy truck GPS trajectory data obtained from the China Road Freight Supervision and Service Platform (https://www.gghypt.net/). The platform provides information for heavy trucks with a maximum load capacity of at least 12 tons.

\subsection{Mobility network construction}\label{mobility-network-construction}

We represent each mobility network as an undirected, weighted network. For the inter-city networks analyzed here, we define TAZs using the administrative boundaries of 336 cities. The origins and destinations of individual trips are assigned to their corresponding cities by determining whether their coordinates fall within the full extent of the administrative urban areas. We then aggregate the flows between different cities and neglect flows whose origin and destination are located within the same city. Through this procedure, we obtain the human mobility flow \(T_{ij}^{H}\) and freight trip flow \(T_{ij}^{F}\) between different cities\(\ i\) and \(j\).

Based on these flows, we construct two undirected, weighted complex networks, i.e., the inter-city human mobility network and the inter-city freight trip network. In both networks, cities are represented as nodes, and the trip flow between cities \(i\) and \(j\) is represented as edge weight \(w_{ij}\).

\subsection{Network measures}\label{network-measures}

\vspace{0.4em}
\noindent\textbf{Edge weight.} 
The edge weight \(w_{ij}\) is the total mobility flow between the connected zones \(i\) and \(j\), aggregated over both directions.

\vspace{0.4em}
\noindent\textbf{Node strength.} 
The node strength \(S_{i}\) is computed as the sum of the weights of all edges incident to node \(i\):
\begin{equation}
\label{eq:node-strength}
S_{i} = \sum_{j \in \Gamma_{i}} w_{ij}
\end{equation}
where \(\Gamma_{i}\) is the set of nodes directly connected to node \(i\).

\vspace{0.4em}
\noindent\textbf{Node disparity.} The node disparity \(Y_{i}\) quantifies the heterogeneity of the weights of edges incident to node \(i\)\cite{ref36}. It is computed as:
\begin{equation}
\label{eq:node-disparity}
Y_{i} = \sum_{j \in \Gamma_{i}} {(w_{ij}/S_{i})}^{2}
\end{equation}
where a low disparity indicates relative uniformity in edge weights, whereas a high \(Y_{i}\) reflects strong heterogeneity.

\vspace{0.4em}
\noindent\textbf{Node weight.} This measure is introduced specifically for the renormalization process. For the original network (\(l = 0\)), node weight coincides with node strength, that is, \(\phi_{i}^{(0)} = S_{i}^{(0)}\). For the renormalized network at layer \(l + 1\), the node weight of a renormalized node \(\mathcal{A}\) is obtained by summing the node weights of all constituent nodes at layer \(l\):
\begin{equation}
\label{eq:node-weight}
\phi_{\mathcal{A}}^{(l + 1)} = \sum_{u \in \mathcal{A}} \phi_{u}^{(l)}
\end{equation}
where \(u \in \mathcal{A}\) indicates that node \(u\) belongs to the renormalized node \(\mathcal{A}\). The node weight thus equals the sum of its external edge weights plus twice the weight of its internal edge weights. Node weight is used solely to rank nodes during box covering and is not involved in subsequent analyses.

\subsection{Neighbor-Limited Box Covering algorithm}\label{neighbor-limited-box-covering-algorithm}

Given a prescribed box mass \(m\), the renormalization of a network then proceeds as follows:

\begin{description}[leftmargin=3.8em, labelwidth=3.2em, labelsep=0.4em, itemsep=0.6em, font=\bfseries]

\item[Step 1:] Sort all nodes in descending order according to their node weight (see the caption of \figref{fig:nlbc-method}) to form an ordered sequence.

\item[Step 2:] Select the node with the highest node weight from the sequence. If this node has at least \(m - 1\) neighbors, group it together with its \(m - 1\) highest-weight neighbors into a box of mass \(m\). Increase the number of boxes \(N(m)\) by one and remove the nodes in this box from the sequence. If the selected node has fewer than \(m - 1\) neighbors, skip it and proceed to the next node in the sequence with at least \(m - 1\) neighbors. Repeat this procedure until no nodes with at least \(m - 1\) neighbors remain.

\item[Step 3:] Apply the same procedure to nodes with at least \(m - 2\) neighbors in the remaining sequence following Step 2. Continue this procedure for nodes with at least \(m - 3,\ m - 4\),\ldots, \(m - (m - 1)\) neighbors, until no more nodes with neighbors remain in the sequence. Then, treat isolated nodes as individual boxes, resulting in a total number of \(N(m)\) boxes.

\item[Step 4:] Merge the nodes within each box into a single renormalized node, whose node weight \(\phi_{\mathcal{A}}^{(l + 1)}\) is computed using \eqnref{eq:node-weight}. The edge weight between renormalized nodes \(\mathcal{A}\) and \(\mathcal{B}\) is computed as:

\begin{equation}
\label{eq:renormalized-edge-weight}
w_{\mathcal{A},\mathcal{B}}^{(l + 1)}
= \sum_{u \in \mathcal{A}}\sum_{v \in \mathcal{B}} w_{u,v}^{(l)}
\end{equation}
where \(w_{u,v}^{(l)}\) is the edge weight between nodes \(u\) and \(v\) at layer \(l\). This yields the next layer of the renormalized network.

\item[Step 5:] The process terminates when only one node remains (see \(l = 2\) in \figref[d]{fig:nlbc-method}); otherwise, return to Step 1.

\end{description}

\subsection{Fractal dimension}\label{fractal-dimension}

The fractal dimension \(d_{f}\)\cite{ref42} of the network is estimated using the box-covering method. In this study, the number of nodes within each box is defined as box mass \(m\). For a given \(m\), the minimum number of boxes required to cover the network is denoted as \(N(m)\), which follows the scaling relation:

\begin{equation}
\label{eq:fractal-dimension}
N(m) \sim m^{- d_{f}}
\end{equation}

The fractal dimension \(d_{f}\) is then obtained from the slope of the linear fit in the log--log plot of \(N(m)\) versus \(m\).

\subsection{Average maximum pointwise deviation}\label{average-maximum-pointwise-deviation}

The maximum pointwise deviation is defined as the largest absolute difference between the curves over all corresponding points. For two curves \(f(x)\) and \(g(x)\) sampled at points \(\{ x_{1},x_{2}\ldots,x_{n}\}\), it is computed as:

\begin{equation}
\label{eq:max-pointwise-deviation}
D_{\max} = \max_{1 < i < n}{\left| f\left( x_{i} \right) - g\left( x_{i} \right) \right|}
\end{equation}

This metric captures the supremum difference between two functions. It retains the key idea of the Kolmogorov--Smirnov statistic, but applying it directly to empirical curves rather than to their cumulative distributions.

The average maximum pointwise deviation is computed as by evaluating \(D_{\max}\) for all pairwise combinations of curves and then taking the mean of these values, denoted as \({\overline{D}}_{\max}\), which quantifies the overall discrepancy within the group.

\subsection{Statistical test for power law}\label{statistical-test-for-power-law}

\noindent\textbf{Null hypothesis \(H_0\):}  The data follow a power-law distribution\(\ Y(x) \sim x^{- \alpha}\);

\noindent\textbf{Alternative hypothesis \(H_1\):} The data do not follow a power-law distribution.

In the statistical test for power law, a large number of power-law distributed synthetic data sets are generated with the same scaling parameter \(\alpha\). Each data set represents one simulated sample. For the empirical data set, the maximum pointwise deviation is computed as:

\begin{equation}
\label{eq:empirical-deviation}
D_{\mathrm{emp}} = \max_{1 < i < n}{\left| Y_{\mathrm{emp}}(x_{i}) - Y_{\alpha}\left( x_{i} \right) \right|}.
\end{equation}

While \(Y_{\mathrm{emp}}(x)\) and \(Y_{\alpha}(x)\) refer to the empirical data points and fitted power-law model. For each synthetic data set, we re-estimate the scaling parameter \(\alpha_{i}\) and compute maximum pointwise deviation between the synthetic data points \(Y_{\mathrm{syn}}(x)\) and fitted power-law model \(Y_{\alpha_{i}}(x)\), i.e.

\begin{equation}
\label{eq:synthetic-deviation}
D_{\mathrm{syn}} = \max_{1 < i < n}{\left| Y_{\mathrm{syn}}(x_{i}) - Y_{\alpha_{i}}(x_{i}) \right|}.
\end{equation}

Finally, the p-value is defined as the proportion of simulations with a larger maximum deviation than the empirical data:

\begin{equation}
\label{eq:p-value}
\text{p-value} = P\left( D_{\mathrm{syn}} \geq D_{\mathrm{emp}} \right)
\end{equation}

A large p-value indicates that the observed discrepancy does not constitute evidence against the model. Conversely, a small p-value suggests that the model does not provide a plausible fit to the data. We use a significance level of 0.1; thus, p-values above 0.1 indicate that the null hypothesis cannot be rejected.

\subsection{Weighted susceptible--infected--susceptible epidemic spreading model}\label{weighted-susceptibleinfectedsusceptible-epidemic-spreading-model}

In this model, each node can exist in one of the two states, i.e., susceptible (\(S\)) or infected (\(I\)), at any given time \(t\). Initially, \(1\%\) of the nodes are randomly selected to be infected, with the remaining nodes susceptible. The original infectivity \(\lambda\) is defined as the infection probability associated with a unit weight (\(w = 1\)).

The epidemic spreads from infected nodes to their neighbors. If the weight between a susceptible node \(i\) and one of its infected neighbors \(j\) is \(w_{ij}\), then node \(j\) infects node \(i\) with probability:

\begin{equation}
\label{eq:weighted-infection-probability}
\lambda\left( w_{ij} \right) = 1 - \left( 1 - \widetilde{\lambda} \right)^{w_{ij}}
\end{equation}
where \(\widetilde{\lambda}\) denotes the rescaled infectivity in the renormalized network. Due to the large disparities in edge weight values among renormalized networks at different scales, we rescale the original infectivity \(\lambda\) to ensure comparability of spreading process across different network layers, inspired by previous studies\cite{ref43}. Specifically, the same original infectivity \(\lambda\) is assigned to each renormalized network, but is rescaled by the average edge weight \(w_{ave}\) of the corresponding network, yielding:

\begin{equation}
\label{eq:rescaled-infectivity}
\widetilde{\lambda} = \lambda/w_{ave}
\end{equation}

This rescaled parameter represents the effective infectivity per unit weight at this given renormalized network. Accordingly, in the SIS spreading process, each renormalized network uses its corresponding effective infectivity \(\widetilde{\lambda}\).

A susceptible node \(i\) becomes infected with probability:

\begin{equation}
\label{eq:susceptible-infection-probability}
1 - \prod_{j \in \Gamma_{i}^{I}} \left[ 1 - \lambda\left( w_{ij} \right)\right]
\end{equation}
where \(\Gamma_{i}^{I}\) is the set of infected neighbors of node \(i\). Infected nodes recover and return to the susceptible state with probability \(\mu = 1\). The spreading process terminates when the number of the infection density stabilizes or a predefined time step is reached.

In this model, whether a node is infected depends solely on the current states of this node and its neighbors, and independent of its previous states.

\subsection{Relative infection proportion}\label{relative-infection-proportion}

For each renormalized network, we compute the relative infection proportion to characterize the weighted SIS infection outcomes, defined as:

\begin{equation}
\label{eq:relative-infection-proportion}
\widetilde{P} = \frac{N_{\lambda_{k}}^{I}}{\sum_{k = 1}^{a}N_{\lambda_{k}}^{I}}
\end{equation}
where \(a\) denotes the number of predefined infectivity values and \(\Delta = \{\lambda_{1},\lambda_{2}\ldots\lambda_{k - 1},\ \lambda_{k}\ldots\ \lambda_{a - 1},\ \lambda_{a}\}\) is the predefined set of infectivity values. Here, \(N_{\lambda_{k}}^{I}\) is the number of infected nodes corresponding to \(k\)-th infectivity \(\lambda_{k}\). The denominator \(\sum_{k = 1}^{a}N_{\lambda_{k}}^{I}\) represents the total number of infected nodes across all predefined infectivity values in \(\Delta\). The relative infection proportion quantity characterizes the relative infection level associated with \(\lambda_{k}\). If the resulting curves of different renormalized networks nearly overlap, this indicates that networks at different scales exhibit consistent relative infection levels for the same infectivity.

\subsection*{Table 1}

\begin{table}[htbp]
\caption{Basic size statistics of renormalized networks (first five layers).}
\label{tab:renormalized-network-size}
\centering
\small
\renewcommand{\arraystretch}{1.6}
\setlength{\tabcolsep}{6pt}

\begin{tabular}{>{\centering\arraybackslash}m{0.16\linewidth}|
                >{\centering\arraybackslash}m{0.17\linewidth}|
                >{\centering\arraybackslash}m{0.07\linewidth}
                >{\centering\arraybackslash}m{0.09\linewidth}
                >{\centering\arraybackslash}m{0.11\linewidth}
                >{\centering\arraybackslash}m{0.11\linewidth}
                >{\centering\arraybackslash}m{0.12\linewidth}}
\hline
Renormalization method & Network type & Layer & Box mass & Number of nodes & Number of edges & Average edge weight \\
\hline

\multirow{10}{*}{LL-Renorm}
& \multirow{5}{*}{\begin{tabular}{c}Inter-city human\\mobility\end{tabular}}
& 0 & 1  & 335 & 30421 & 157 \\
& & 1 & 2  & 171 & 10764 & 395 \\
& & 2 & 4  & 88  & 3311  & 1120 \\
& & 3 & 8  & 46  & 926   & 3332 \\
& & 4 & 16 & 25  & 265   & 10264 \\
\cline{2-7}
& \multirow{5}{*}{\begin{tabular}{c}Inter-city freight\\trips\end{tabular}}
& 0 & 1  & 336 & 32860 & 426 \\
& & 1 & 2  & 171 & 10892 & 997 \\
& & 2 & 4  & 88  & 3279  & 2599 \\
& & 3 & 8  & 46  & 929   & 6928 \\
& & 4 & 16 & 24  & 258   & 18869 \\
\hline

Renormalization method & Network type & Layer & Box mass & Number of nodes & Number of edges & Average edge weight \\
\hline

\multirow{10}{*}{SL-Renorm}
& \multirow{5}{*}{\begin{tabular}{c}Inter-city human\\mobility\end{tabular}}
& 0 & 1 & 335 & 30421 & 157 \\
& & 1 & 2 & 171 & 10764 & 395 \\
& & 1 & 3 & 116 & 5482  & 726 \\
& & 1 & 4 & 87  & 3304  & 1170 \\
& & 1 & 5 & 74  & 2331  & 1664 \\
\cline{2-7}
& \multirow{5}{*}{\begin{tabular}{c}Inter-city freight\\trips\end{tabular}}
& 0 & 1 & 336 & 32860 & 426 \\
& & 1 & 2 & 171 & 10892 & 997 \\
& & 1 & 3 & 115 & 5420  & 1835 \\
& & 1 & 4 & 87  & 3268  & 2762 \\
& & 1 & 5 & 70  & 2172  & 3984 \\
\hline
\end{tabular}
\end{table}

\section*{Acknowledgements}

This work was supported by the National Natural Science Foundation of China (Grant Nos. 72288101, 72271019, 72242102).

\clearpage

\begin{center}
{\Large\bfseries Supplementary Information\par}
\end{center}

\vspace{1em}

\captionsetup[figure]{
  justification=raggedright,
  singlelinecheck=false
}

\setcounter{figure}{0}
\renewcommand{\thefigure}{S\arabic{figure}}
\setcounter{table}{0}
\renewcommand{\thetable}{S\arabic{table}}
\setcounter{equation}{0}
\renewcommand{\theequation}{S\arabic{equation}}

\section{Intra-city Data Description}\label{sec:supp-intra-city-data-description}

With rapid urbanization, over half of the world's population lives in cities, making both inter-city and intra-city human mobility increasingly crucial\cite{ref1}. Similarly, inter-city and intra-city freight mobility are also essential for modern economic development. They facilitate the flow of goods and resources between regions, maintaining a balance between supply and demand. Consequently, while the manuscript examines the multi-scale characteristics of inter-city human and freight mobility, the supplementary information focuses on the multi-scale characteristics of real-world intra-city human and freight mobility networks.

Here, four Chinese cities are selected as the study areas. Human mobility networks are constructed within the urban built-up areas, where high human population density regions are predominantly concentrated. In contrast, freight trip networks are constructed within the urban municipal districts. In many Chinese cities, industrial parks and large warehouses are mainly located on the periphery of built-up areas, resulting in intensive freight activities in these districts. Figure S1 illustrates the spatial distributions of human population density and heavy truck outflows in the four cities. It shows that urban built-up areas largely encompass regions with high population density, but do not fully cover areas with intensive freight activity. Accordingly, we consider urban built-up areas to be more appropriate for constructing intra-city human mobility networks, whereas municipal districts provide a more suitable spatial extent for intra-city freight mobility networks.

\begin{figure}[!htbp]
\centering
\includegraphics[width=0.95\linewidth,height=0.72\textheight,keepaspectratio]{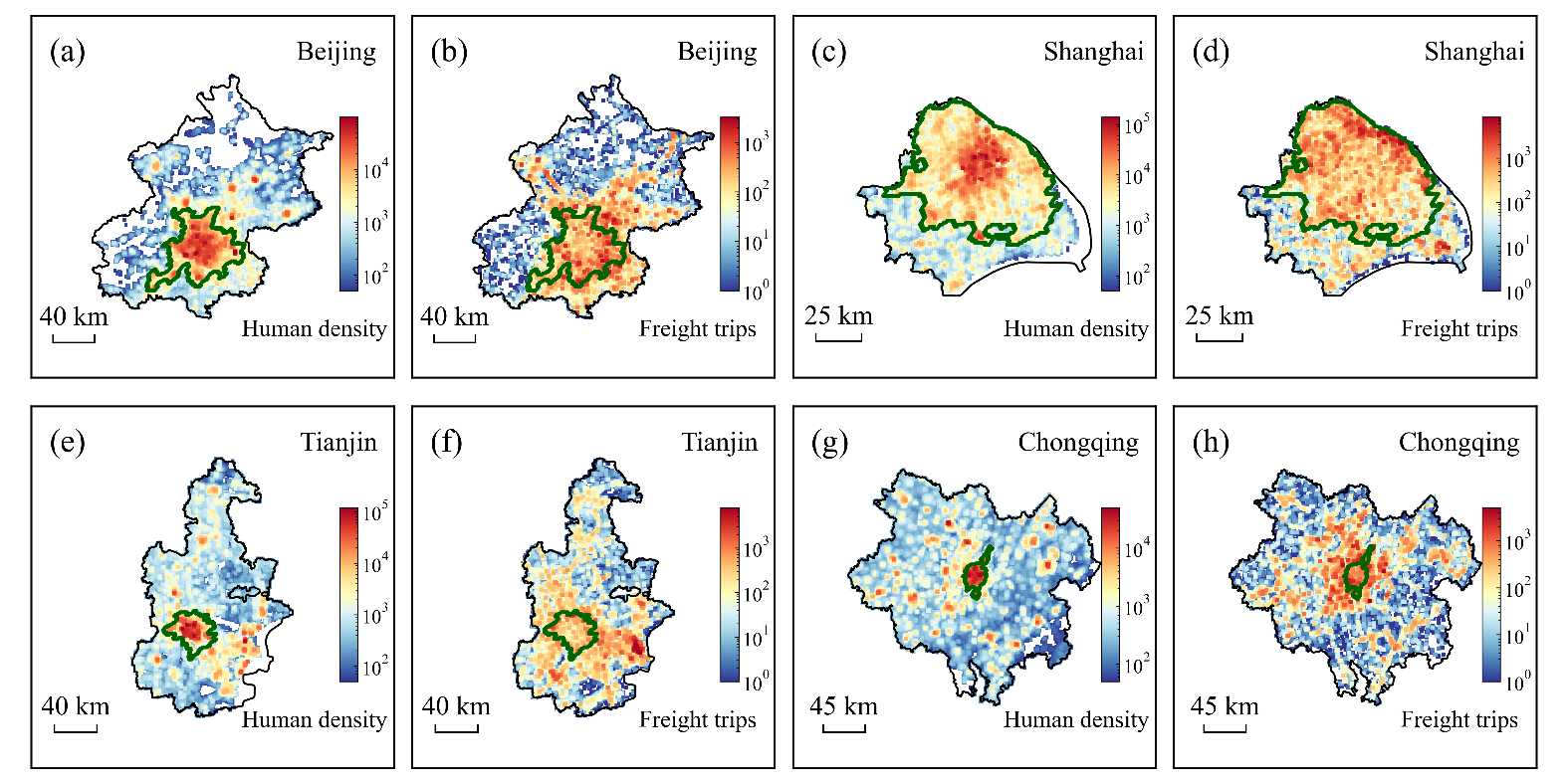}
\caption*{\textbf{Fig. S1} Spatial distributions of human population density and heavy truck trip outflows in (a-b) Beijing, (c-d) Shanghai, (e-f) Tianjin and (g-h) Chongqing. Black lines represent the urban municipal district boundaries, and green lines indicate urban built-up area boundaries. The color of each unit indicates the human population density and number of heavy truck trip outflows, respectively. Urban built-up area data is obtained from ``Description of the GHS Urban Centre Database 2015''. Human population density data is obtained from \url{https://www.worldpop.org/}.}
\label{fig:supp-s1}
\end{figure}

\section{Data processing and network construction}\label{sec:supp-data-processing-and-network-construction}

The intra-city human mobility dataset is provided by a Chinese telecommunications operator. It consists of non-personally identifiable movement records representing approximately 90 million residents across 60 cities during August and November 2019. Each trip record contains the date, hour, and geographic coordinates (longitude and latitude) of both the origin and destination. The intra-city heavy truck trip data is from Yang et al. (2022), consistent with those used in the main manuscript.

Intra-city mobility networks are constructed following the same procedure as described in the main manuscript. Traffic analysis zones (TAZs) are defined as spatial units of approximately 1.27 km\textsuperscript{2}, corresponding to Google S2 cells \url{https://github.com/google/s2geometry}. The origins and destinations of individual trips are mapped to their corresponding spatial units. We then aggregate the flows between different TAZs and neglect flows whose origin and destination are located within the same TAZ. Through this procedure, we obtain the human mobility flow \(T_{ij}^{H}\) and freight trip flow \(T_{ij}^{F}\) between different TAZs \(i\) and \(j\).

Based on these flows, the intra-city human mobility networks and the intra-city freight trip networks of the four Chinese cities are constructed as undirected, weighted networks (see Fig. S2). In these networks, each TAZ is represented as a node, and edges are weighted according to the mobility flows between the TAZs \(i\) and \(j\). Table S1 shows the size of the intra-city mobility networks in terms of the numbers of nodes and edges.

\begin{figure}[!htbp]
\centering
\includegraphics[width=0.95\linewidth,height=0.72\textheight,keepaspectratio]{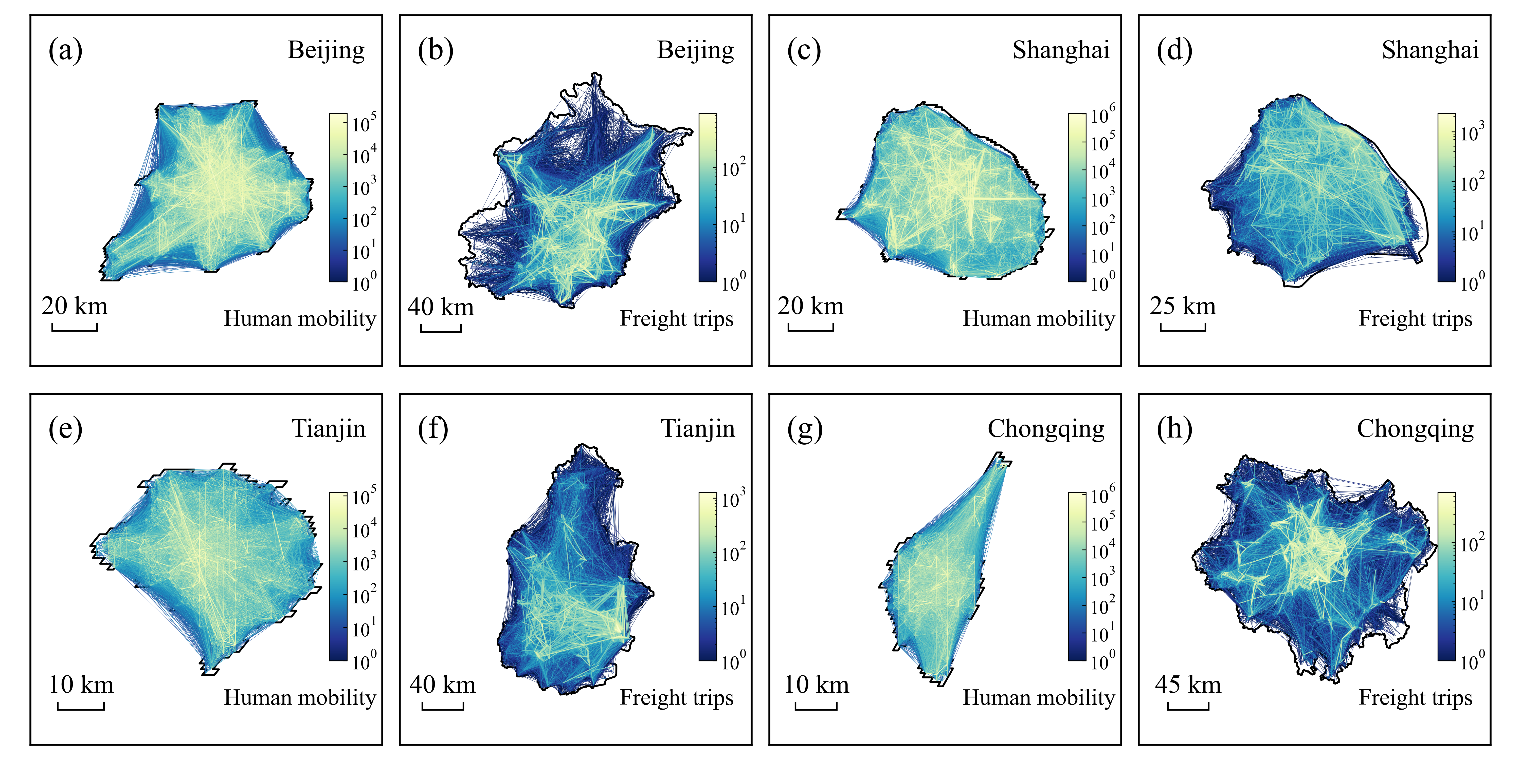}
\caption*{\textbf{Fig. S2} Expected desire line maps of intra-city human mobility and freight trip networks in (a-b) Beijing, (c-d) Shanghai, (e-f) Tianjin and (g-h) Chongqing. Here, human mobility networks are mapped onto urban built-up areas, whereas freight trip networks are mapped onto urban municipal districts.}
\label{fig:supp-s2}
\end{figure}

\begin{table}[!htbp]
\centering
\caption*{\textbf{Table S1}\\
Basic statistics for intra-city mobility networks.}
\label{tab:supp-s1}
\renewcommand{\arraystretch}{1.35}
\begin{tabular*}{0.95\linewidth}{@{\extracolsep{\fill}}cccc}
\toprule
City & Network Type & Number of Nodes & Number of Edges \\
\midrule
\multirow{2}{*}{Beijing} & Human & 1,884 & 1,131,594 \\
& Freight & 8,726 & 231,004 \\
\midrule
\multirow{2}{*}{Shanghai} & Human & 1,514 & 809,115 \\
& Freight & 4,422 & 330,436 \\
\midrule
\multirow{2}{*}{Tianjin} & Human & 613 & 144,944 \\
& Freight & 9,211 & 257,546 \\
\midrule
\multirow{2}{*}{Chongqing} & Human & 337 & 49,487 \\
& Freight & 11,729 & 232,533 \\
\bottomrule
\end{tabular*}
\end{table}

\section{Self-similarity of multi-scale intra-city mobility networks}\label{sec:supp-self-similarity-intra-city}

We apply LL-Renorm and SL-Renorm to the intra-city human mobility and freight trip networks of Beijing, Shanghai, Tianjin, and Chongqing. Under each renormalization scheme, we estimate the fractal dimension of the resulting networks to quantify their topological similarity. Figures S3.1--S3.4 present the log-log plots of the box number \(N(m)\) versus box mass \emph{m}, indicating an approximately power-law dependence on \emph{m} within a finite range of scales. To determine the valid scaling range for the power-law fitting, we adopt a p-value-based statistical criterion. Specifically, starting from the smallest box scale, we progressively extend the fitting window toward larger scales and perform a goodness-of-fit test at each step. The upper bound of the power-law regime is identified as the largest scale for which the fit satisfies \(p > 0.1\). As shown in Fig. S3.1--S3.4, power-law behavior is most evident in the early stages of renormalization. Although deviations emerge at larger scales, a stable scaling relationship \(N(m)\) and \emph{m} persists over a finite range of smaller scales. These findings indicate that both LL-Renorm and SL-Renorm effectively capture the self-similar, multi-scale structure of urban mobility networks, supporting their applicability as a general framework for the structural analysis in large-scale complex networks.

\begin{figure}[!htbp]
\centering
\includegraphics[width=0.95\linewidth,height=0.72\textheight,keepaspectratio]{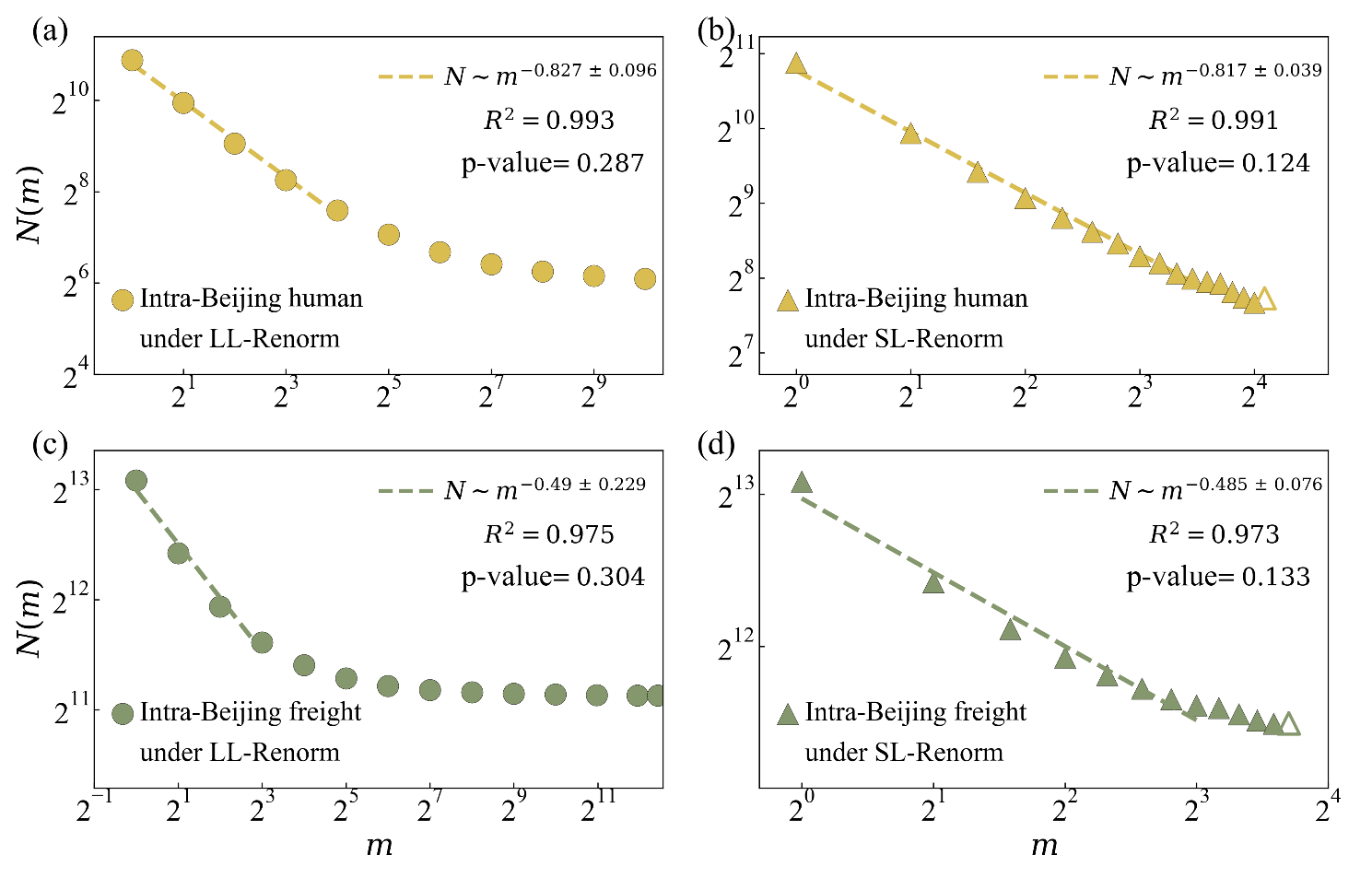}
\caption*{\textbf{Fig. S3.1} Self-similar scaling in intra-Beijing mobility networks.}
\label{fig:supp-s3-1}
\end{figure}

\begin{figure}[!htbp]
\centering
\includegraphics[width=0.95\linewidth,height=0.72\textheight,keepaspectratio]{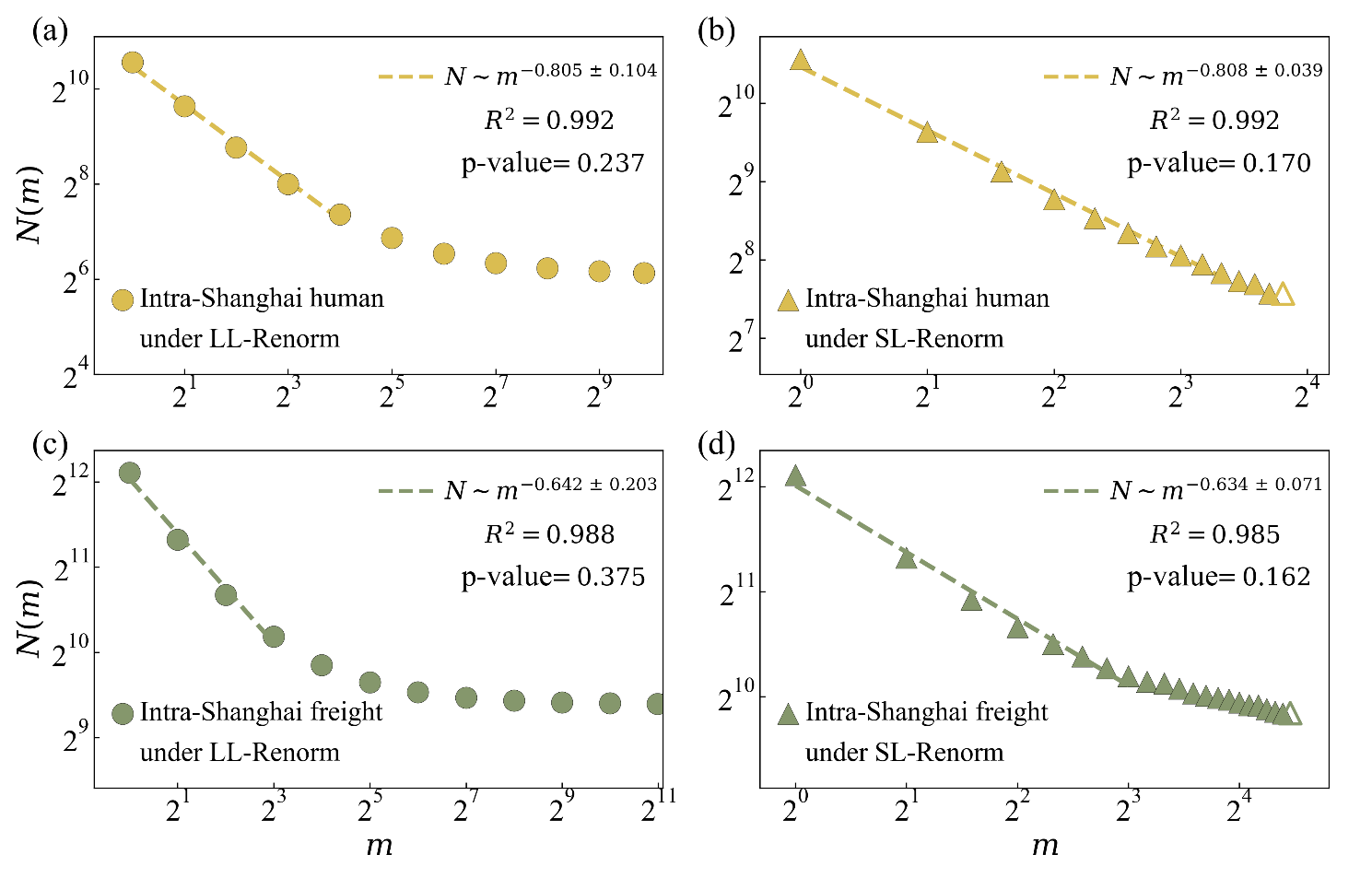}
\caption*{\textbf{Fig. S3.2} Self-similar scaling in intra-Shanghai mobility networks.}
\label{fig:supp-s3-2}
\end{figure}

\begin{figure}[!htbp]
\centering
\includegraphics[width=0.95\linewidth,height=0.72\textheight,keepaspectratio]{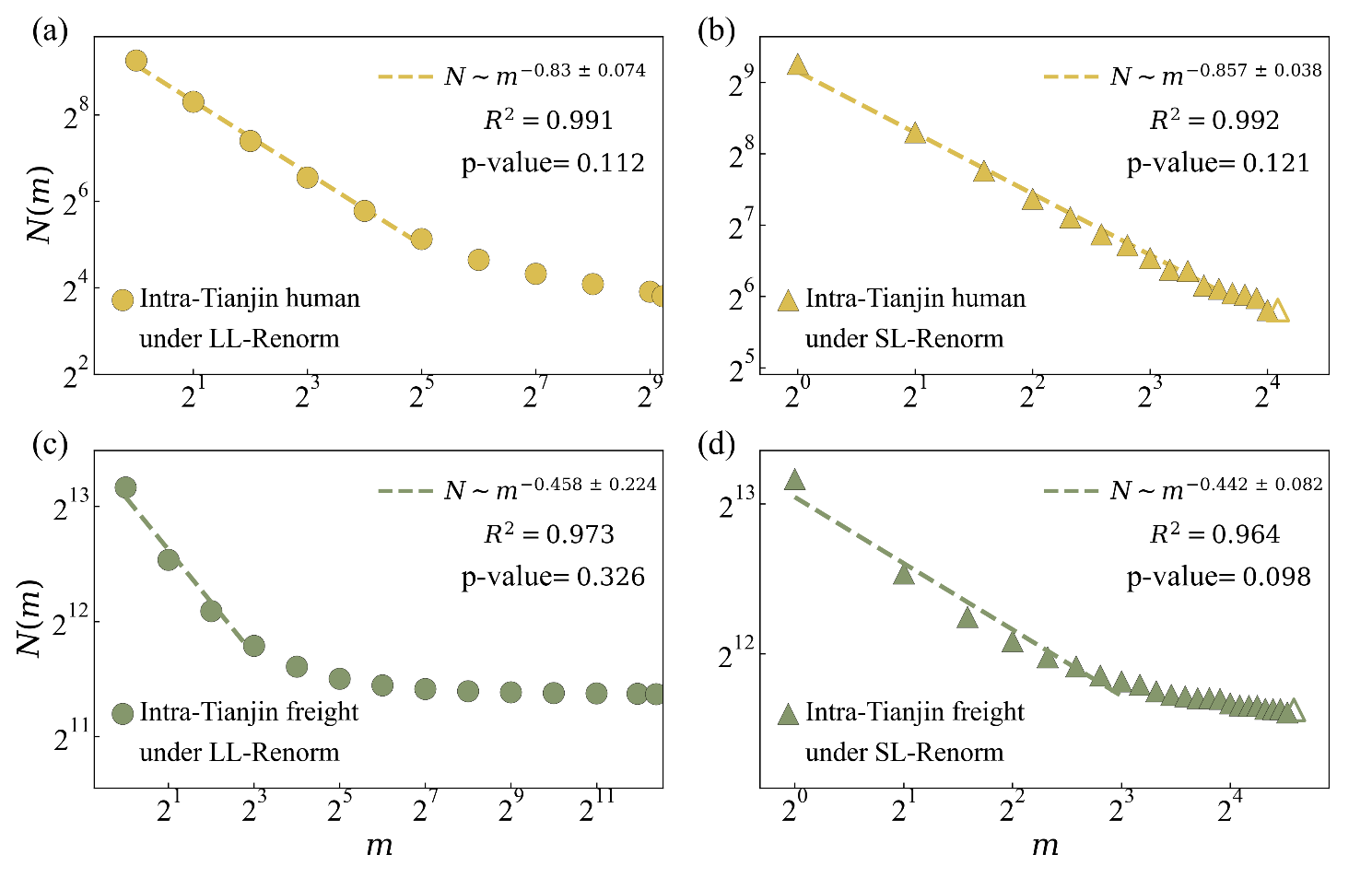}
\caption*{\textbf{Fig. S3.3} Self-similar scaling in intra-Tianjin mobility networks.}
\label{fig:supp-s3-3}
\end{figure}

\begin{figure}[!htbp]
\centering
\includegraphics[width=0.95\linewidth,height=0.72\textheight,keepaspectratio]{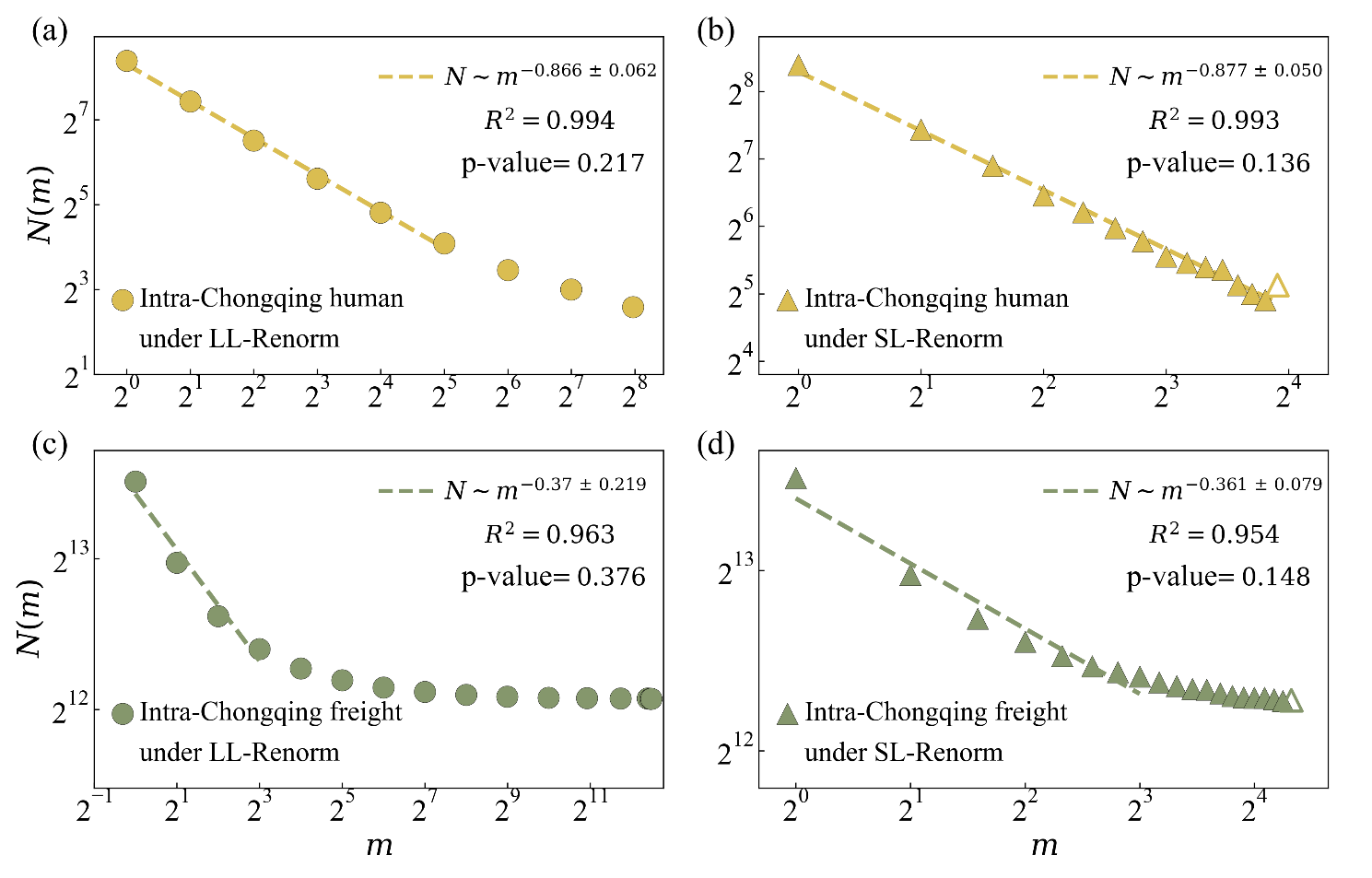}
\caption*{\textbf{Fig. S3.4} Self-similar scaling in intra-Chongqing mobility networks.}
\label{fig:supp-s3-4}
\end{figure}

We further analyze the edge weight \(w_{ij}\), node strength \(S_{i}\), and node disparity \(Y_{i}\) across multi-scale intra-city mobility networks to assess the similarity of their weighted structural properties. As shown in Fig. S4.1-S4.4, the complementary cumulative distribution functions (CCDFs) of edge weights, node strengths, and node disparities for the four cities almost merge onto nearly identical curves after rescaling by their respective layer-averaged values. This similarity is further quantified using the maximum pointwise deviation \({\overline{D}}_{\max}\) across the set of CCDFs. These results demonstrate that the weighted structural features of both intra-city human mobility and freight trip networks exhibit statistical self-similarity under LL-Renorm and SL-Renorm transformations.

\begin{figure}[!htbp]
\centering
\includegraphics[width=0.92\linewidth,height=0.78\textheight,keepaspectratio]{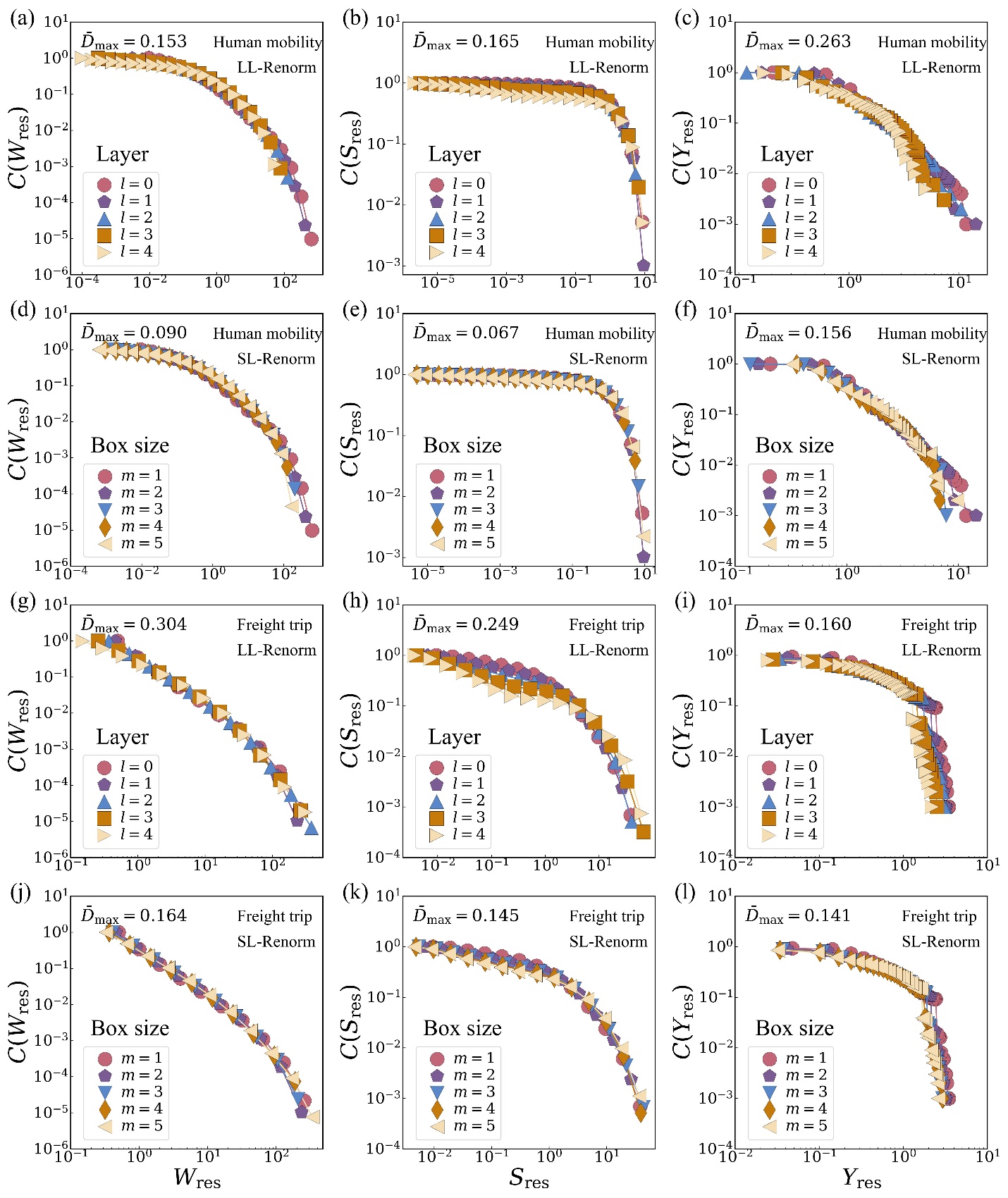}
\caption*{\textbf{Fig. S4.1} The weighted structural features for the multi-scale intra-Beijing mobility networks.}
\label{fig:supp-s4-1}
\end{figure}

\begin{figure}[!htbp]
\centering
\includegraphics[width=0.92\linewidth,height=0.78\textheight,keepaspectratio]{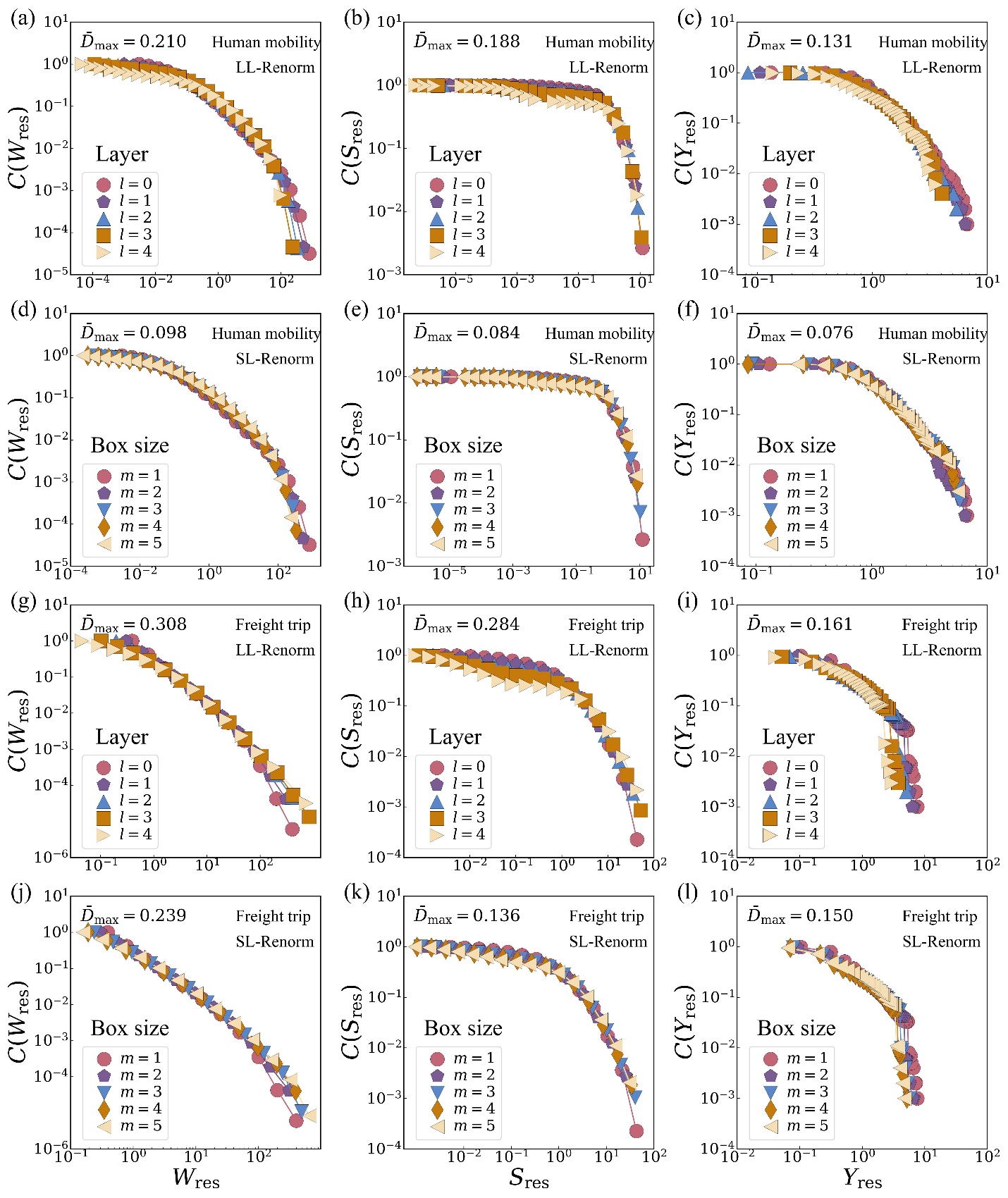}
\caption*{\textbf{Fig. S4.2} The weighted structural features for the multi-scale intra-Shanghai mobility networks.}
\label{fig:supp-s4-2}
\end{figure}

\begin{figure}[!htbp]
\centering
\includegraphics[width=0.92\linewidth,height=0.78\textheight,keepaspectratio]{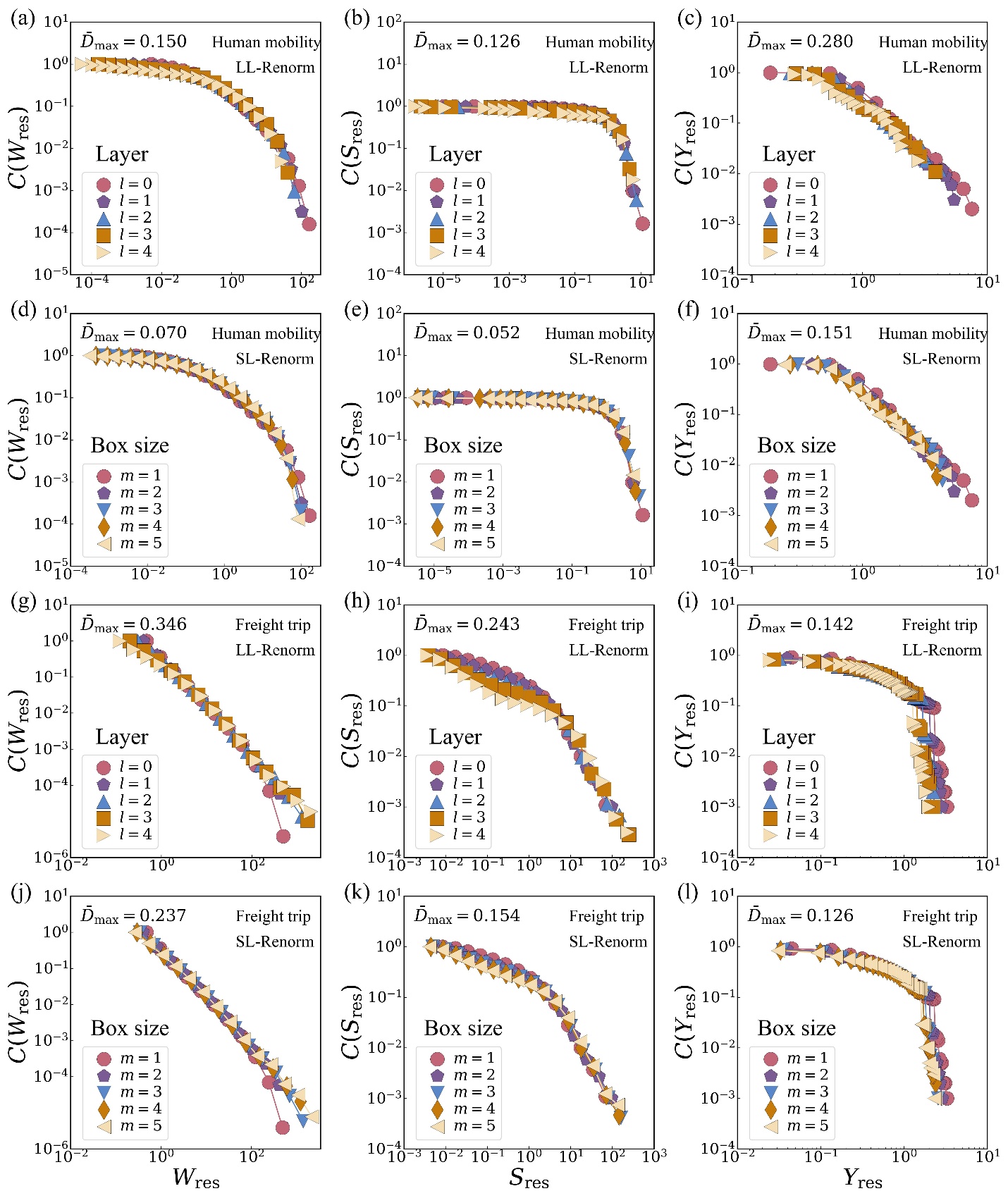}
\caption*{\textbf{Fig. S4.3} The weighted structural features for the multi-scale intra-Tianjin mobility networks.}
\label{fig:supp-s4-3}
\end{figure}

\begin{figure}[!htbp]
\centering
\includegraphics[width=0.92\linewidth,height=0.78\textheight,keepaspectratio]{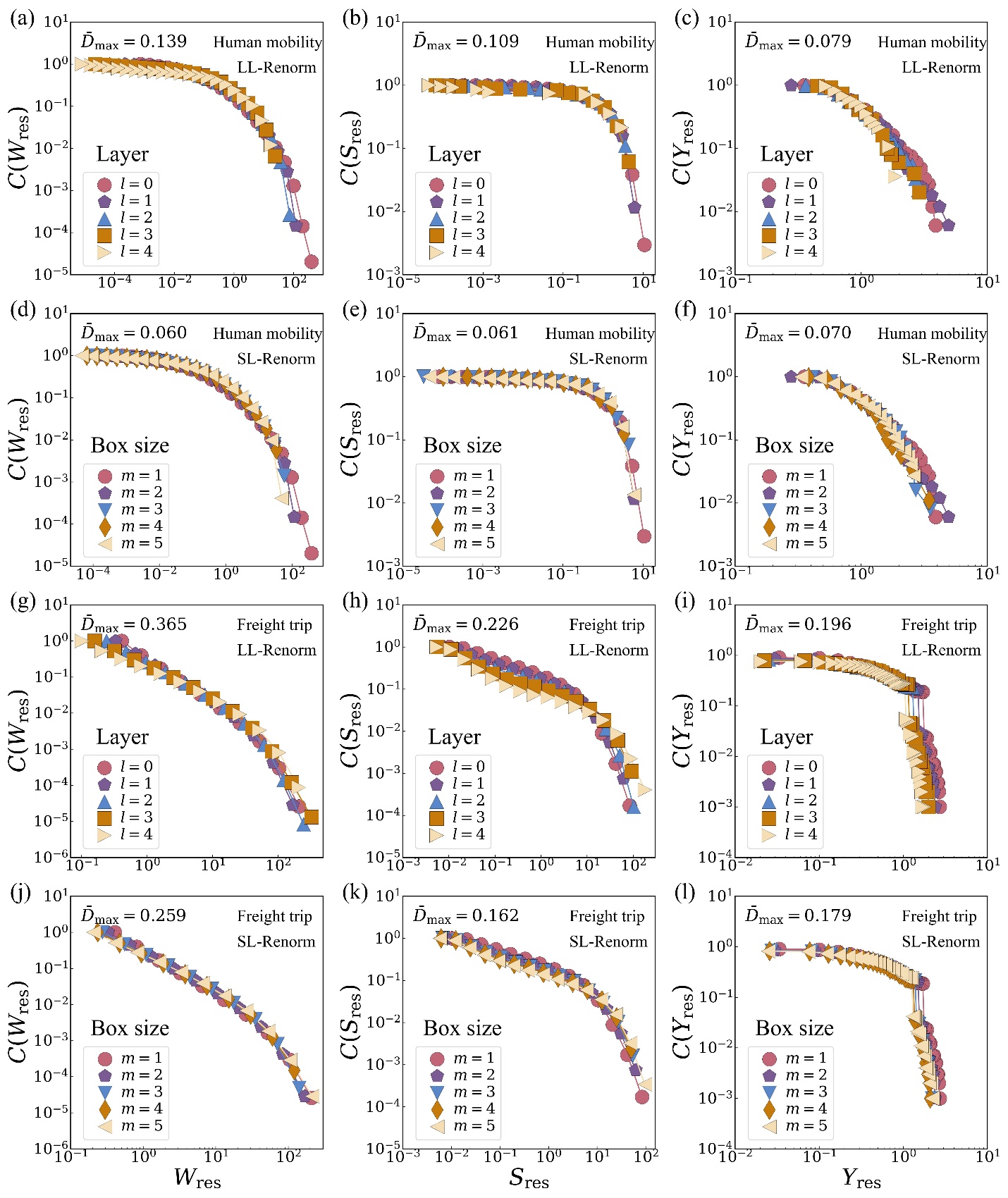}
\caption*{\textbf{Fig. S4.4} The weighted structural features for the multi-scale intra-Chongqing mobility networks.}
\label{fig:supp-s4-4}
\end{figure}

\end{document}